\documentclass[pra,a4paper,showpacs,twocolumn,superscriptaddress,longbibliography]{revtex4-1}

\usepackage{amssymb}
\usepackage{amsmath}
\usepackage{amsfonts}
\usepackage{graphicx}
\usepackage{bm}
\usepackage{color}
\usepackage{multirow}
\usepackage{natbib}
\usepackage{hyperref}
\usepackage[normalem]{ulem}
\usepackage{relsize}

\DeclareMathOperator{\tr}{tr}

\DeclareMathOperator{\im}{Im}
\DeclareMathOperator{\re}{Re}
\DeclareMathOperator{\erf}{erf}
\DeclareMathOperator{\erfi}{erfi}

\def\br{\bm{r}}

\newcommand{\corr}[1]{\langle #1\rangle}
\newcommand{\ccorr}[1]{\langle\langle #1\rangle\rangle}

\def\be{\begin{equation}}
\def\ee{\end{equation}}

\begin{document}

\title{Magnetic disorder in superconductors: Enhancement by mesoscopic fluctuations}

\author{I. S.~Burmistrov}

\affiliation{Skolkovo Institute of Science and Technology, 143026 Moscow, Russia}

\affiliation{L. D. Landau Institute for Theoretical Physics, Kosygina
  street 2, 117940 Moscow, Russia}

\affiliation{Laboratory for Condensed Matter Physics, National Research University Higher School of Economics, 101000 Moscow, Russia}

\author{M. A. Skvortsov}

\affiliation{Skolkovo Institute of Science and Technology, 143026 Moscow, Russia}

\affiliation{L. D. Landau Institute for Theoretical Physics, Kosygina
  street 2, 117940 Moscow, Russia}

\date{\today} 

\begin{abstract}
We study the density of states (DOS) and  the transition temperature $T_c$ in a dirty superconducting film with rare classical magnetic impurities
of an arbitrary strength described by the Poissonian statistics. We take into account that the potential disorder is a source for mesoscopic fluctuations of the local DOS, and, consequently, for the effective strength of magnetic impurities. We find that these mesoscopic fluctuations result in a non-zero DOS for all energies in the region of the phase diagram where without this effect the DOS is zero within the standard mean-field theory. This mechanism can be more efficient in filling the mean-field superconducting gap than rare fluctuations of the potential disorder (instantons). Depending on the magnetic impurity strength, the suppression of $T_c$ by spin-flip scattering can be faster or slower than in the standard mean-field theory.
\end{abstract}

\maketitle

\section{Introduction}

The properties of superconductors in the presence of impurities have remained at the focus of intense theoretical and experimental research during the past half-century.
It is generally accepted that the potential scattering in $s$-wave superconductors
affects neither the transition temperature, $T_c$, nor the density of sta\-tes (DOS),
$\rho(E)$. This statement usually referred to as Anderson's theorem \cite{AG0a,AG0b,Anderson1959} is valid for sufficiently good metals.
As the potential disorder increases, the
emergent inhomogeneity due to the interplay of quantum inter\-ference (Anderson localization) and interaction leads to modification of $T_c$ \cite{Maekawa1982,Maekawa1984,Muttalib1983,Fin1987,Feigelman2007,Feigelman2010,FS2012,BGM2012,BGM2015} and $\rho(E)$ \cite{Maki1970,Abrahams1970,Castro1990,BGM2016},
with the ef\-fect being controlled by the parameter $1/(k_F l) \ll 1$ (where $k_F$ is the Fermi momentum and $l$ is the mean free path).

\begin{figure}
\centerline{\includegraphics[width=0.99\columnwidth]{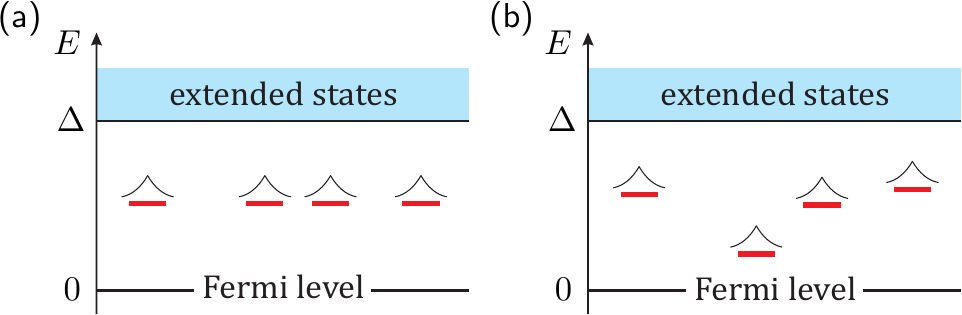}\hspace{0.02\textwidth}}
\caption{(Color online) Subgap states localized at individual magnetic impurities. (a) In a clean system, the energies of all bound states are equal.
(b) Mesoscopic fluctuations lead to the log-normal distribution of impurity strength [cf.\ Eq.\ \eqref{eq:Pdef}], rendering the energies of bound states position-dependent.}
\label{figure-levels}
\end{figure}

Magnetic impurities violating the time-reversal symmetry affect superconductivity much stronger, already at $k_F l\to \infty$.
Classical magnetic impurities lead both to suppression of $T_c$ and to reduction of the superconducting gap in $\rho(E)$ with the increase of their concentration $n_s$ \cite{AG2}. Beyond the Born limit, magnetic impurities produce degenerate subgap bound states (see Fig.\ \ref{figure-levels}a). Their hybridization results in the formation of an energy band giving rise to a nontrivial DOS structure \cite{Yu,Soda,Shiba,Rusinov}. The account for the Kondo effect \cite{MHZ1971,Matsuura1977,Bickers1987}, the indirect exchange interaction between magnetic impurities
\cite{Ruvalds1981}, or the spin-flip scattering assisted by the electron-phonon interaction \cite{Jarrell1990} can lead to the reentrant behavior of $T_c$ vs.\ $n_s$ (see Ref.\ \cite{Balatsky} for a review).

A hard gap in $\rho(E)$ obtained for superconductors with magnetic impurities in the mean-field approximation is smeared by inhomogeneity. This can be due to rare fluctuations of a potential disorder \cite{LamS-1,LamS-2,MS,MaS},
$n_s$ \cite{SilvaIoffe}, or superconducting order parameter \cite{LO}.
A combined theory of these mechanisms has been developed in Refs.\ \cite{SkF,FS2016}.

In this Letter we describe a novel mechanism for smearing of the superconducting gap. We reconsider the problem of rare classical magnetic impurities with the Poissonian statistics in a dirty superconductor. The key point that distinguishes our work from the previous ones is that we take into account mesoscopic fluctuations of the local DOS in a potential disorder. Physically, this implies that the energies of subgap bound states become dependent on the spatial positions of magnetic impurities (see Fig.\ \ref{figure-levels}b).
Averaging over these bound states results in a non-zero homogenous DOS
at all energies in the region of the phase diagram
where in the absence of this effect $\rho(E)$ is zero within the mean-field approximation.
Motivated by the recent experiment on magnetic Gd impurities in superconducting MoGe films \cite{Rogachev2012}, in this Letter we develop the theory of the enhancement of magnetic disorder by mesoscopic fluctuations in the case of a dirty superconducting film.

The outline of the paper is as follows. In Sec.\ \ref{Sec3} we present description of dirty superconductors with rare magnetic impurities in terms of the nonlinear sigma model and its renormalization. Our results for the renormalized spin-flip rate, superconducting transition temperature, and the density of states are given in Sec.\ \ref{Sec4}. We end the paper with discussions (Sec.\ \ref{Sec5}) and conclusions (Sec.\ \ref{Sec6})
Some details of calculations are presented in Appendices. 

\section{Nonlinear sigma model  for paramagnetic impurities\label{Sec3}}

We consider a two-dimensional (2D) dirty $s$-wave superconductor in the presence of both potential (spin-preserving) and magnetic disorder. Scattering off the former is responsible for the dominant contribution to the momentum relaxation rate $1/\tau$.
Much weaker spin-flip scattering rate is related with the exchange interaction between magnetic impurities and electrons described by the Hamiltonian
\begin{equation}
{H}_{\rm mag} = J \sum_{j} \psi^\dag(\bm{r}_j) \bm{S}_j \bm{\sigma} \psi(\bm{r}_j)  .
\end{equation}
We shall treat rare magnetic disorder under standard assumptions \cite{Yu,Soda,Shiba,Rusinov,MaS}:
(i) impurity positions $\bm{r}_j$ have the Poisson distribution; (ii) impurity spins $\bm{S}_j$ are classical statistically independent vectors with the flat distribution over their orientations, $\prod_j \delta (\bm{S}_j^2-S^2)$.

The low energy description of two-dimensional disordered superconductors with rare paramagnetic impurities can be conveniently formulated in terms of the replicated verion of a nonlinear sigma-model \cite{Fin,BK}.
Its action can be written as 
\begin{equation}
\label{SSS}
  \mathcal S =  \mathcal S_D + \mathcal S_\Delta + \mathcal S_\text{mag} .
\end{equation}
Here $\mathcal S_D$ is the standard diffusive action
\begin{equation}
\label{S0}
\mathcal S_D
= \frac{\pi\nu}8 \int d^2 \bm{r} \tr \bigl[ D (\nabla Q)^2 - 4 (\varepsilon \tau_3 + \Delta \tau_1) Q \bigr] ,
\end{equation}
where $\nu$ and $D$ denote the density of states at the Fermi energy (per one spin projection) and the diffusive coefficient in the normal state, respectively. The matrix $Q$ operates in the spin, Nambu, replica, and Matsubara energy spaces. It is subject to the following constraints~\cite{Houzet_Skvortsov}:
\be
\label{Q-constraint}
Q^2=1,\qquad   Q = \overline Q \equiv \tau_1 \sigma_2 Q^\textsf{T} \tau_1 \sigma_2 .
\ee
Here the transposition ${}^\textsf{T}$ acts in both the Matsubara energy space and the replica space. The Pauli matrices $\tau_j$ ($\sigma_j$)
act in the Nambu (spin) spaces. The matrix $\varepsilon$ is the diagonal matrix with the elements $\varepsilon_n= \pi T(2n+1)$.

The superconducting correlations are described by the order-parameter matrix $\Delta$ which is diagonal in the Nambu space with matrix elements $\Delta^a(\bm{r})$. In the absence of a supercurrent, $\Delta$ is chosen to be real. The action $\mathcal S_\Delta$ reads
\begin{equation}
\mathcal S_\Delta = \frac{\nu}{\lambda T} \int d^2\bm{r} \sum_{a=1}^N |\Delta^a(\bm{r})|^2 .
\end{equation}
Here $N$ stands for the number of replica and $\lambda>0$ denotes the attraction amplitude in the Cooper-channel.

We consider the case of rare classical magnetic impurities with the concentration $n_s$
[the precise condition on $n_s$ see below], when
the magnetic part of the action, $\mathcal S_\text{mag}$, becomes separable in the individual magnetic impurities~\cite{MS}:
\begin{equation}
\mathcal S_\text{mag} \approx \sum_j s_\text{mag}^{(j)} = -\frac 12 \sum_j \tr\nolimits \ln \left( 1+ i\sqrt{\alpha} \, Q(\bm{r}_j)\tau_3 \bm{\sigma} \bm{n}_j \right) .
\label{Ssep}
\end{equation}
Here $\bm{n}_j$ stands for the three-dimensional unit vector and the dimensionless parameter $\alpha=(\pi \nu JS)^2$ is expressed in terms of the impurity spin $S$ and exchange constant $J$. We note that approximation \eqref{Ssep} of the full action $\mathcal S_\text{mag}$
is equivalent to the self-consistent $T$-matrix approximation for magnetic scattering which treats all orders in scattering off a single magnetic impurity but neglects diagrams with intersecting impurity lines.

Performing the Poisson averaging over positions of the magnetic impurities with the help of the following relation~\cite{Friedberg1975}
\begin{equation}
\Bigl \langle \exp \sum_j f(\bm{r_j}) \Bigr \rangle = \exp \left \{ n_s \int d^2\bm{r} \Bigl [e^{f(\bm{r})}-1\Bigr ]\right \} ,
\end{equation}
we find that the contribution to the nonlinear sigma model action due to magnetic impurities becomes
\begin{equation}
\mathcal S_\text{mag} \to
    - {n}_s \int d^2 \bm{r}
  \left ( \left \langle e^{\frac 12 \tr\nolimits \ln \left( 1+ i\sqrt{\alpha} \, Q(\bm{r})\tau_3 \bm{\sigma} \bm{n} \right)}\right \rangle_{\bm{n}}
  -1 \right) .
\label{Smagn0}
\end{equation}
Here $\langle \dots \rangle_{\bm{n}}$ stands for the averaging over direction of the unit vector $\bm{n}$.
Expanding $\mathcal{S}_\text{mag}$ in powers of $\sqrt{\alpha}$, we find
\begin{align}
  \mathcal{S}_\text{mag}
 = &
  - {n}_s \int d^2\bm{r}
  \Biggl[
    \sum_{m=1}^\infty \frac{(-1)^{m-1}}{2m}
      \hat C T_m \notag \\
 & + \frac{1}{2!} \sum_{m,n=1}^\infty \frac{(-1)^{m+n}}{4mn}
      \hat C T_{mn}
      \notag \\
     &   + \frac{1}{3!} \sum_{m,n,p=1}^\infty \frac{(-1)^{m+n+p-1}}{8mnp}
      \hat C T_{mnp}
  + \dots
  \Biggr] .
\end{align}
Here we introduced the operators
\begin{align}
  \hat C T_m
   = {} &
  C_{i_1\dots i_m} \tr (Q {A}_{i_1} \dots Q {A}_{i_m})
, \notag 
\\   
\hat C T_{mn}
   = {} & 
  C_{i_1\dots i_{m+n}}
  \tr (Q {A}_{i_1} \dots Q {A}_{i_m}) \notag 
\\
  & \times 
  \tr (Q {A}_{i_{m+1}} \dots Q {A}_{i_{m+n}})  ,
\end{align}
and so on. The operator $\hat C$ acts as the symmetric tensor:
$C_{i_1\dots i_m}
  =
  \corr{n_{i_1} \dots n_{i_m}}$ .
For convenience we defined the self-dual matrix
$
\bm{A} = i\sqrt{\alpha} \tau_3 \bm{\sigma}
= \overline{\bm{A}}$ .
Since operators $T_{n\dots}$ are symmetric with respect to its indices, the expansion can be written in the following form:
\begin{align}
  \mathcal{S}_\text{mag}
  = &
 {n}_s \int d^2\bm{r}
  \Biggl[
    \frac{1}{4} \hat CT_2
  - \frac{1}{8} \hat CT_{11}
  + \frac{1}{8} \hat CT_4
  - \frac{1}{12} \hat CT_{31}
  \notag\\
  & - \frac{1}{32} \hat CT_{22}
  + \frac{1}{32} \hat CT_{211}
  - \frac{1}{384} \hat CT_{1111}
 + \dots
  \Biggr] .
  \label{Smag:full}
\end{align}

The nonlinear sigma model action \eqref{SSS} with the magnetic part given by Eq. \eqref{Smag:full} provides full description of quantum effects for a dirty superconductor in the diffusive regime. These effects (weak localization and Aronov-Altshuler-type corrections) are responsible for the renormalization of system's parameters, e.g. the diffusion coefficient and the attraction amplitude. In the 2D case, the magnitude of quantum corrections at the energy scale $\varepsilon$
is governed by the parameter
\begin{equation}
\label{eq:t(E)}
  t(\varepsilon) = \frac{1}{\pi g} \ln \frac{1}{|\varepsilon|\tau}
,
\end{equation}
where $g = h/(e^2 R_\square) \gg 1$ is the bare dimensionless conductance of the film. In a superconductor, renormalization stops at $\varepsilon\sim\max\{T_c, |\Delta|\}\sim T_c$.
Assuming that the transition temperature is not too low, $t(T_c)\ll1$, one can neglect the renormalization of the conductance and interaction parameters between the energy scales $1/\tau$ and $T_c$ (see Refs.\ \cite{Fin,BK} for a review).
In contrast, renormalization of the magnetic-impurity part $S_{\rm mag}$ of the nonlinear sigma model is essential.

Treating this renormalization in the one-loop approximation, we find that after the renormalization this part of the action can be written as (see Appendix \ref{App1})
\begin{gather}
  \mathcal S_\text{mag}
  = 
  n_s \int d^2\br
  \Biggl[
    \gamma_2 \hat CT_2
  + \gamma_{11} \hat CT_{11}
  + \gamma_4 \hat CT_4
  + \gamma_{31} \hat CT_{31}\notag \\
   + \gamma_{22} \hat CT_{22}  + \gamma_{211} \hat CT_{211}
  + \gamma_{1111} \hat CT_{1111}
  + \dots
  \Biggr] .
  \label{Smag-ser}
\end{gather}
Here the coefficients $\gamma_{k_1k_2\dots k_q}$, where $k_1+k_2+\dots+k_q= n$, are given as follows:
\be
  \gamma_{k_1k_2\dots k_q}(t)
  =
  \gamma_{k_1k_2\dots k_q}(0) e^{n(n-1)t} ,
  \label{eq:multi-spectrum}
\ee
where initial values of the coefficients $\gamma_{k_1k_2\dots k_q}(0)$ follow from Eq. \eqref{Smag:full}.

In what follows we are interested in the singlet sector of the theory. Therefore, one can operate with $Q$ matrix which is the unit matrix in the spin space, $Q=Q_0 \sigma_0$. Then we can average over directions of the impurity magnetization $\bm{n}$ in operators $\hat C T_{k_1k_2\dots k_q}$. 
Then the renormalized magnetic-impurity part of the nonlinear sigma-model action \eqref{Smag-ser} can be written in the following convenient short-hand notation (see Appendix \ref{App2}):
\be
\label{Smag-full}
 \mathcal S_\text{mag}
  =
  - {n}_s \int d^2\bm{r}
  \left \langle
  \exp\left\{
    \frac14 \tr \ln[1+ \textsf{a} (Q \tau_3)^{2}]
  \right\}
  - 1
  \right \rangle_\textsf{a} ,
\ee
where the averaging $\langle \dots \rangle_\textsf{a}$  is defined with respect to the following log-normal distribution function:
\begin{equation}
\mathcal{P}_\alpha(\textsf{a},t) = \frac{1}{4 \textsf{a}\sqrt{\pi {t}}} \exp \left [ -\frac{1}{4 t} \left ( \frac{1}{2}\ln \frac{\textsf{a}}{\alpha}+t\right )^2\right ] .
\label{eq:Pdef}
\end{equation}

Comparing Eq.\ \eqref{Smag-full} with Eq.\ \eqref{Smagn0} we may interpret the effect of renormalization as follows: Now instead of a single value of $\alpha$ there is a log-normal distribution of the effective strength of impurity $\textsf{a}$, schematically shown in Fig.\ \ref{figure-levels}.

\section{Results\label{Sec4}}

In the mean-field approximation, a dirty superconductor in the diffusive regime is described by two coupled equations: the self-consistency equation for the superconducting order parameter $\Delta$ and the Usadel equation for the quasiclassical Green's function \cite{Usadel,Beltzig,Kopnin}. These equations can be derived as the saddle-point equations of the nonlinear sigma model described in the previous section. The mean-field solution for the $Q$ matrix can be parametrized as
\be
Q=\tau_1\sin\theta+\tau_3\cos\theta .
\label{eq:Q:conf}
\ee
Performing variation of the action $\mathcal{S}$ on the configuration \eqref{eq:Q:conf} with respect to $\Delta$, we find the following mean-field self-consistency equation:
\begin{equation}
\Delta = \pi \lambda T \sum_\varepsilon \sin \theta_\varepsilon .
\label{eq:Supp:SCE}
\end{equation}

For the study of space-averaged configurations at the mean-field level, it is sufficient to retain the term of the first order in trace only in the renormalized action  (\ref{Smag-full}):
\be
 \mathcal S_\text{mag}^\text{MF}
  =
  - \frac{n_s}{4} \int d^2\bm{r}
  \left\langle
   \tr \ln[1+ \textsf{a} (Q \tau_3)^{2}]
  \right \rangle_\textsf{a} .
  \label{eq:action:MF}
\ee
Since the eigenvalues of $(Q\tau_3)^2$ are $e^{\pm2i\theta}$, we find explicitly
\be
  \mathcal S_\text{mag}^\text{MF}
  =
-  \frac{n_s}{2} \sum\limits_{\sigma=\pm} \int d^2\br
  \left \langle
     \ln\bigl (
    1 + \textsf{a}\, e^{2i\sigma\theta}
  \bigl  ) \right \rangle_\textsf{a} .
   \label{eq:action:MF2}
\ee
Performing variation of the action $\mathcal{S}$ [with $\mathcal S_\text{mag}$ given by Eq.~\eqref{eq:action:MF2}] on the configuration \eqref{eq:Q:conf} with respect to $\theta_\varepsilon$, we find the following modified Usadel equation:
\begin{equation}
\varepsilon \sin \theta_\varepsilon - \Delta \cos \theta_\varepsilon + \frac{n_s}{\pi \nu} \left \langle \frac{\textsf{a} \sin2{\theta}_\varepsilon}{1+\textsf{a}^2+2\textsf{a} \cos2\theta_\varepsilon} \right \rangle_\textsf{a} = 0
,
\label{eq:MF-UE}
\end{equation}
where $\theta_\varepsilon$ is the energy-dependent spectral angle, $\nu$ is the normal DOS at the Fermi energy per
one spin projection, and $\varepsilon= \pi T (2 n+1)$ denotes the fermionic Matsubara frequencies. We remind that the averaging $\langle \dots \rangle_\textsf{a}$  in Eq. \eqref{eq:MF-UE} is defined with respect to the log-normal distribution function \eqref{eq:Pdef}.
Since $\mathcal{P}_\alpha(\textsf{a},t\to 0)  \to \delta(\textsf{a}-\alpha)$, Eq.~\eqref{eq:MF-UE} at $t=0$ coincides with the standard Usadel equation in the case of magnetic impurities \cite{AG2,Shiba,Rusinov,MaS}.
The linearity of Eq.~\eqref{eq:MF-UE} in $n_s$ is justified for small concentration of magnetic impurities:
$n_s \xi^2/g\ll 1$, where $\xi=l/\sqrt{T_c\tau}$ is the dirty superconducting coherence length.

The quantity $\textsf{a}$ in Eq.\ \eqref{eq:MF-UE} plays a role of the renormalized
impurity strength. Since the bare impurity strength $\alpha$
is proportional to the local DOS which is subjected to mesoscopic fluctuations,
$\mathcal{P}_\alpha(\textsf{a},t)$
reflects the log-normal distribution of the local DOS in 2D weakly disordered systems \cite{AKL1986,Lerner1988}. Contrary to naive expectations, one should average over $\textsf{a}$ the Usadel equation rather than physical observables, e.g.\ the DOS. This is a consequence of the Poisson distribution of impurity positions $\bm{r}_j$.

\begin{figure}
\centerline{\includegraphics[width=0.95\columnwidth]{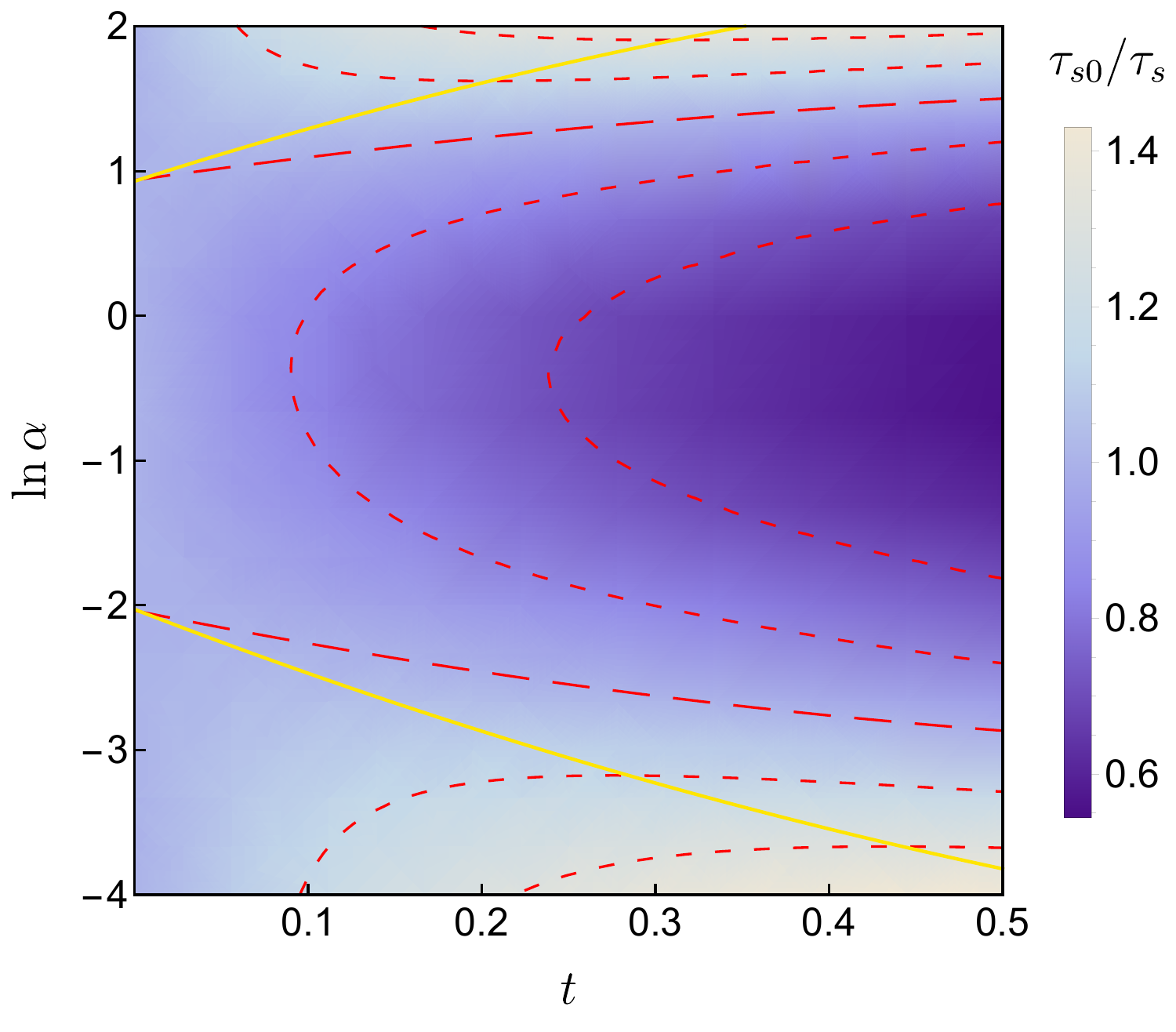}}
 \caption{(Color online) The color plot for $\tau_{s0}/\tau_s$ vs.\ $\ln \alpha$ and $t$.
The red dashed curves indicate isolines 0.7, 0.85, 1 (long dash), 1.15, 1.3.
The yellow curves mark the position of the maximum of $\tau_{s0}/\tau_s$ as a function of $t$ for a fixed value of $\alpha$.
}
 \label{figure-tau}
\end{figure}

\subsection{Effective spin-flip rate}

In the vicinity of the thermal transition, $\Delta\to 0$ and we can linearize Eq.\ \eqref{eq:MF-UE} with respect to $\theta_\varepsilon$. This procedure yields  
\begin{equation}
\theta_\varepsilon \approx \Delta/(\varepsilon+ 1/\tau_s) ,
\label{eq:Delta:lin}
\end{equation}
where the effective spin-flip rate is given by
\begin{equation}
\frac{1}{\tau_s} = \frac{2 n_s}{\pi \nu} \left \langle \frac{\textsf{a}}{(1+\textsf{a} )^2} \right \rangle_\textsf{a}  .
\label{eq:SFR}
\end{equation}
At $t=0$, one recovers the standard expression for the bare spin-flip rate due to magnetic impurities, $1/\tau_{s0} = 2 \alpha n_s/[\pi \nu (1+\alpha)^2]$ \cite{Rusinov}.
In the limiting cases $\alpha \to 0$ and $\alpha\to\infty$, the spin-flip rate \eqref{eq:SFR} becomes enhanced in comparison with the bare one: $1/\tau_s = \exp(2 t) /\tau_{s0}$ and $1/\tau_s = \exp(6 t) /\tau_{s0}$, respectively.
For an arbitrary value of $\alpha$, the asymptotic expansion at $t\ll 1$ has the form (see Appendix \ref{App3:1}):
\begin{equation}
\frac{\tau_{s0}}{\tau_s} \approx 1 + \frac{2-16\alpha+6\alpha^2}{(1+\alpha)^2} t(T_c) + O(t^2) .
\label{eq:tau-ren}
\end{equation}
At small $t$, the spin-flip rate is suppressed (enhanced) for $\alpha_0<\alpha<1/(3\alpha_0)$ (otherwise), where $\alpha_0 =1/(4+\sqrt{13})\approx 0.13$. The overall behavior of the ratio $\tau_{s0}/\tau_s$ as a function of $t$ and $\alpha$ is illustrated in Fig.\ \ref{figure-tau}.

\begin{figure}
\centerline{\includegraphics[width=0.95\columnwidth]{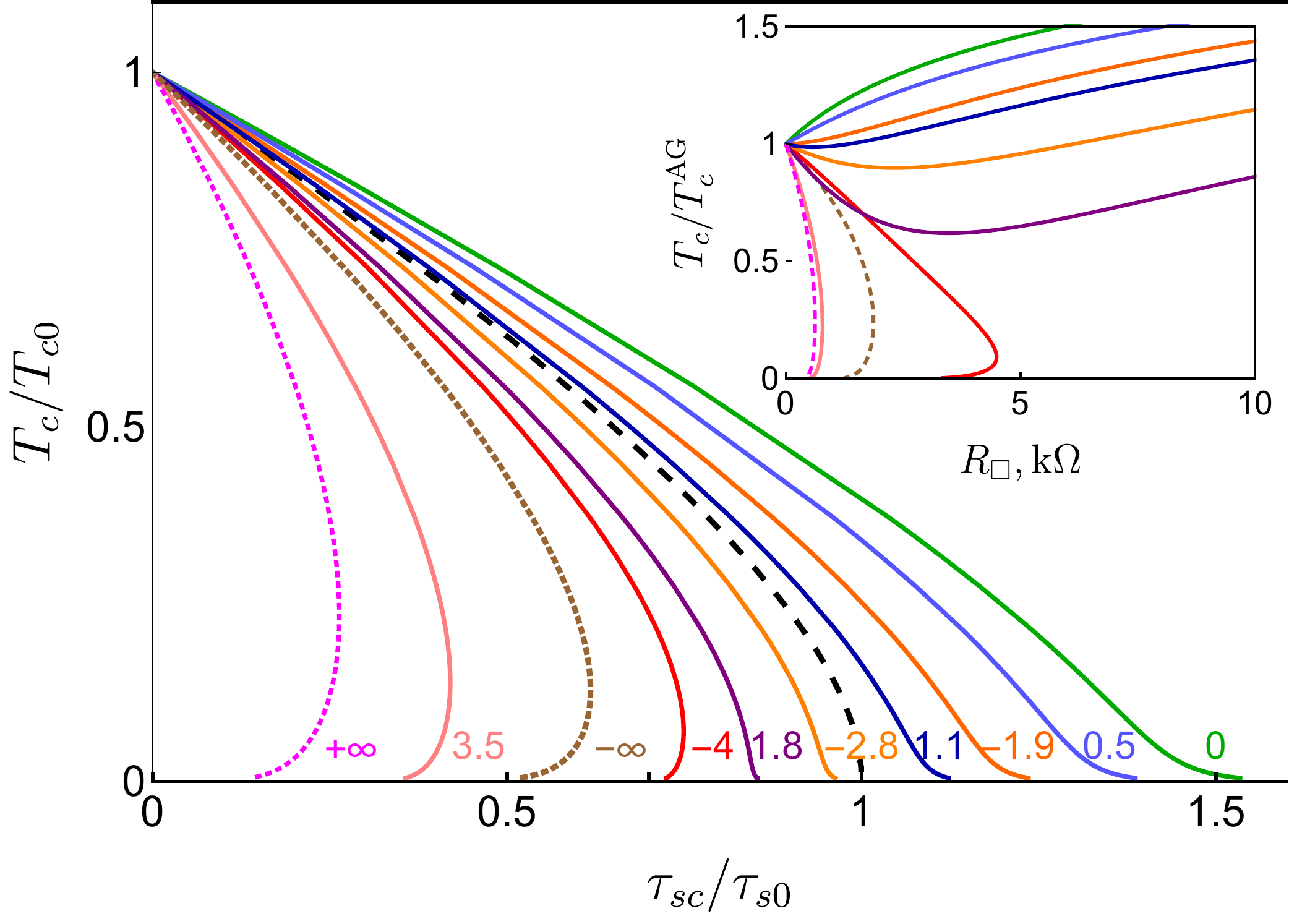}}
 \caption{(Color online) The dependence of $T_c/T_{c0}$ on the bare spin-flip rate $1/\tau_{s0}$ for some values of the bare impurity strength (values of $\ln \alpha$ are indicated near the curves), and $R_\square = 2.5$\! k$\Omega$ ($g=10$). The black dotted curve, $T_{c}^\text{AG}(1/\tau_{s0})$, is the solution of Eq.\ \eqref{eq:MFTc} without renormalization.
Inset: The dependence of $T_c/T_{c}^\text{AG}$ on $R_\square$ for the same values of $\ln \alpha$, and
$\tau_{sc}/\tau_{s0} = 0.7$. We use $\ln (T_{c0}\tau) = 5$.}
 \label{figure-Tc}
\end{figure}

\subsection{Transition temperature}

Since the spin-flip rate (\ref{eq:SFR}) is the only characteristic of magnetic disorder that enters the linearized solution for $\theta_\varepsilon$, we obtain the standard equation for the superconducting transition temperature (see Appendix \ref{App3:2}):
\begin{equation}
\ln \frac{T_{c0}}{T_c} = \psi \left (\frac{1}{2} + \frac{1}{2\pi T_c \tau_s} \right ) - \psi \left (\frac{1}{2} \right ) ,
\label{eq:MFTc}
\end{equation}
where $T_{c0}$ denotes the transition temperature in the absence of magnetic impurities, and $\psi(z)$ stands for the di-gamma function. Equation \eqref{eq:MFTc} was derived by Abrikosov and Gor'kov 
(AG)
in the Born limit ($\alpha\to0$) \cite{AG2}, and later was shown to describe the suppression of $T_c$ for arbitrary values of $\alpha$ \cite{Rusinov}. For a scale-independent spin-flip time, $\tau_s=\tau_{s0}$, Eq.~\eqref{eq:MFTc} defines a universal function $T_c^\text{AG}(1/\tau_{s0})$ shown by the black dashed line in Fig.\ \ref{figure-Tc}. Superconductivity is eventually destroyed at the critical spin-flip rate $1/\tau_{sc}=2\pi e^{\psi(1/2)} T_{c0} \approx 0.88 \, T_{c0}$ \cite{AG2}.
This standard approach corresponds to the limit $t=0$, when mesoscopic fluctuations can be neglected.

An essential modification introduced by the log-normal distribution of the impurity strength (\ref{eq:Pdef}) is that now the spin-flip rate $1/\tau_s$ depends on the parameter $t(T_c)$, i.e.\ on the conductance $g$ and the transition temperature $T_c$ itself. This leads to an unusual behavior illustrated in Fig.~\ref{figure-Tc}, where we present the numerical solutions of Eq.~\eqref{eq:MFTc} for fixed values of $g$ and $T_{c0}\tau$ and
for various values of~$\alpha$.
At finite $t$, dependence of $1/\tau_s$ on $T_c$ renders the curves $T_c(1/\tau_{s0})$ sensitive to a particular value of $\alpha$.
In the range $\alpha_0<\alpha < 1/(3\alpha_0)$, the spin-flip rate decreases monotonously down to zero with increasing $t$. Therefore the reduction of $T_c$ with the increase of $1/\tau_{s0}$ is slower than for $t=0$. This agrees qualitatively with the slowdown of $T_c$ suppression with increasing the film resistance measured in Ref.\ \cite{Rogachev2012}.
In the opposite case, for $\alpha<\alpha_0$ and $\alpha>1/(3\alpha_0)$, the dependence of $T_c$ on $1/\tau_{s0}$ is qualitatively different since the ratio $\tau_{s0}/\tau_s$ can be larger than unity and is a non-monotonous function of $t$. Since the spin-flip rate is enhanced, the reduction of $T_c$ with the increase of $1/\tau_{s0}$ is faster than in the case $t=0$. The non-monotonicity of $\tau_{s0}/\tau_s$ results in the existence of two solutions of Eq.\ \eqref{eq:MFTc} for $T_c$.
Formally, 
it
admits the solution with nonzero $T_c$ for any value of the parameter $\tau_{sc}/\tau_{s0}$. However, we remind that our approach is valid provided the inequality $T_c \gg \exp(-\pi g)/\tau$ holds.

The dependence of the spin-flip rate on $g$ transforms into the dependence of $T_c$ on the film conductance. To illustrate this effect, we fix the value of the parameter $\tau_{sc}/\tau_{s0}$ and plot the ratio $T_c/T_c^\text{AG}(1/\tau_{s0})$ on the film resistance $R_\square$ for some values of $\alpha$ in the inset to Fig.\ \ref{figure-Tc}. Since  for $\alpha_0<\alpha<1/(3\alpha_0)$ the spin-flip rate decreases monotonously with the increase of $t$,  $T_c$ is enhanced with respect to $T_c^\text{AG}$. The non-monotonous dependence of $1/\tau_s$ on $t$ obtained for $\alpha<\alpha_0$ and $\alpha>1/(3\alpha_0)$ leads to the reentrant behavior of $T_c$ on $R_\square$.

It is worthwhile to mention that not only
the suppression of $T_c$ by magnetic impurities but also the reduction of  $\Delta$ is modified at finite $g$ due to the log-normal distribution of the effective impurity strength \cite{elsewhere}.

\subsection{Density of states}

Consider now the superconducting phase with a finite $\Delta$.
The DOS can be obtained from the solution
of Eq.\ \eqref{eq:MF-UE} after analytic con\-ti\-nua\-tion
to real energies $E$:
$\rho(E) = 2\nu \re \cos \theta_{-iE+0}$.
It is convenient to parametrize the spectral angle as $\theta=\pi/2+i \psi$.
Without renormalization ($t=0$), the angle $\psi(E)$ should be determined from equation $F_E(\psi)=0$, where \cite{AG2,Shiba,Rusinov,MaS}
\begin{equation}
F_E(\psi) = \sinh \psi -\frac{E}{\Delta} \cosh\psi -
\frac{[\alpha n_s/(\pi \nu \Delta)]\sinh 2\psi}{1+\alpha^2 -2\alpha \cosh 2 \psi}.
\label{eq:MF-UE-t0}
\end{equation}
This leads to a complicated structure of the DOS at energies $|E|<\Delta$, which depends on the values of $\alpha$ and $\eta=1/\tau_{s0}\Delta$ (see Ref.\ \cite{FS2016} for a review). In the case $\eta > (\frac{1-\alpha}{1+\alpha})^2$, the impurity band touches the Fermi energy, leading to a finite DOS at $E=0$. 
Below we shall consider the opposite regime, $\eta < (\frac{1-\alpha}{1+\alpha})^2$, in which $\rho(E)$ has a finite gap $E_{g0}$ for $t=0$.
The gap opens since $F_E(\psi)=0$ possesses only real solutions at energies $|E|<  E_{g0}$.

\begin{figure}
\centerline{\includegraphics[width=0.95\columnwidth]{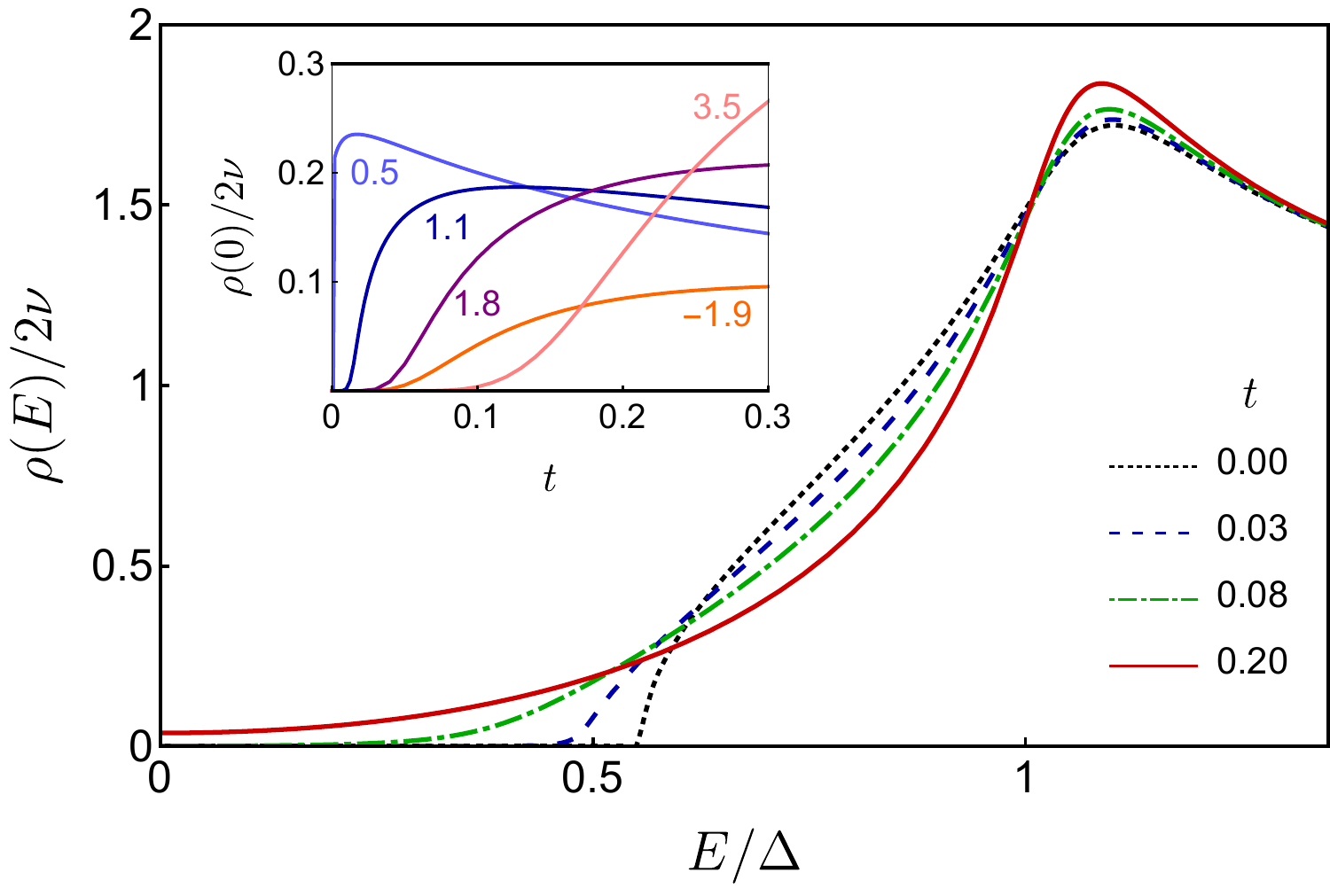}}
\caption{(Color online)
The energy dependence of the DOS for some values of the parameter $t$, and $\alpha=0.05$.
Inset: The DOS at the Fermi energy as a function of $t$ for values of $\ln \alpha$  indicated near the curves.
We use $1/(\tau_{s0}\Delta) = 0.1$.}
 \label{figure-Rho-E}
\end{figure}

Typical modification of the DOS at finite $t$ is illustrated in Fig.~\ref{figure-Rho-E}, where we plot $\rho(E)$ obtained by numerical solution of Eq.\ \eqref{eq:MF-UE} at $\varepsilon\to-iE+0$.
For $t\ll1$,
mesoscopic fluctuations of magnetic disorder affect the DOS in two ways.
(i) On the \emph{perturbative}\ level, they shift the position of the gap:
$E_{g0}\to E_g$, with $(E_{g0}-E_g)/E_{g0} \propto t$, but the gap remains hard. 
(ii) A finite DOS below the renormalized gap is then generated \emph{nonperturbatively}\ in $t$, due to the tail of the distribution $\mathcal{P}_\alpha(\textsf{a},t)$.
In Fig.~\ref{figure-Rho-E}, the smearing of $E_g$ can be clearly seen for $t=0.03$, whereas for larger $t$ the smearing and gap shift cannot be separated.
A profound feature of the DOS is its finite value right at the Fermi energy. 




In the limit of weak renormalization, $t\ll1$, the DOS can be obtained analytically.
The general expression is quite cumbersome (see Appendix \ref{App3:3}), so we present here only the results in the regime of weak magnetic impurities ($\alpha\ll\eta^{2/3}\ll1$).
The gap smearing at $E\to E_g$ is described by
\begin{equation}
\label{dos-near}
\rho(E)/2\nu = \sqrt{2/3}\,\eta^{-2/3}\, {\textrm{Re}\,} \sqrt{\epsilon+i \epsilon_*} ,
\end{equation}
where 
$\epsilon = (E-E_g)/\Delta$
and
\begin{equation}
\epsilon_* = \frac{\eta^{2/3}\sqrt{2\pi}}{16 \sqrt{t}}\left (\frac{\eta^{2/3}}{\alpha}\right )^{3/4} \exp \left ( -\frac{1}{16 t}\ln^2\frac{\eta^{2/3}}{4\alpha} \right ) .
\end{equation}
The subgap DOS (\ref{dos-near}) decays as power law.
The residual DOS at the Fermi energy it is determined by the probability
$\mathcal{P}_\alpha(1,t)$ to find $\textsf{a}=1$ and in the limit $t\ll1$ reads
\begin{equation}
\frac{\rho(0)}{2\nu} 
= 
\frac{\sqrt{\pi}\eta}{8\alpha^{3/4}\sqrt{t}} \exp \left( -\frac{1}{16 t} \ln^2 \frac{1}{\alpha} \right) .
\label{e:RHO-E0}
\end{equation}
This result
is non-perturbative in both $t$ and $\alpha$. 
The dependence of $\rho(0)$ on $t$ for some values of $\ln\alpha$
is shown in the inset to Fig.~\ref{figure-Rho-E}. 
Its non-monotonicity is related to that of  $\mathcal{P}_\alpha(1,t)$ as a function of $t$.
 At a fixed value of $t$, $\rho(0)$ behaves non-monotonically with the impurity strength $\alpha$ at a given value of $\eta$.

\section{Discussions\label{Sec5}}

Our main Eq.~\eqref{eq:MF-UE} could be derived for a toy model of Poissonian magnetic impurities with the strength independently distributed according to $\mathcal{P}_\alpha(\textsf{a},t)$. We emphasize however that in a disordered film the log-normal distribution is generated intrinsically due to mesoscopic fluctuations of the local DOS.

The log-normal distribution  $\mathcal{P}_\alpha(\textsf{a},t)$ predicts an exponentially small probability for realization of very small and very large values of the effective impurity strength~$\textsf{a}$. As well-known from the theory of mesoscopic fluctuations of the local DOS and wave function multifractality, this implies that typically the impurity strength $\textsf{a}_-<\textsf{a} < \textsf{a}_+$ is realized \cite{Mirlin2000}. Using results of Ref.\ \cite{Foster2009}, we obtain the following estimate for the termination points: $\textsf{a}_\pm =\alpha \exp[\pm (4/\sqrt{\pi g}) \ln 1/(T_c\tau)]$ (see Appendix \ref{App4}). In order our result \eqref{e:RHO-E0} for $\rho(0)$ were applicable to a typical sample, vicinity of $\textsf{a}=1$ should be inside the interval $(\textsf{a}_-,\textsf{a}_+)$. It is fulfilled provided $ (4/\sqrt{\pi g}) \ln 1/(T_c\tau) \gg \ln(1/\alpha)$.

We emphasize the difference with the instanton analysis \cite{MaS,FS2016}, where the effect of mesoscopic fluctuations on magnetic disorder was not taken into account: 
(i) in our approach, the DOS is modified already at the mean-field level and (ii) our results \eqref{dos-near}--\eqref{e:RHO-E0} involve a \emph{spreading resistance}\ $\ln(\xi/l)/2\pi g$ which is parametrically larger than the sheet resistance $1/g$ emerging in the instanton analysis. As a result, our mechanism predicts a larger DOS at the Fermi level, whereas near $E_g$ it prevails provided 
$\ln^2 [\eta^{2/3}/(4\alpha)] \ll 16 t\ln [g \eta^{1/2}/(\alpha^{3/4}\sqrt t)]$
(see Appendix \ref{App3:3}).

Although our results were derived for a weak disorder, $t \ll 1$, they can be extended to the case of a moderate disorder, $t \sim 1$ (provided $g\gg 1$) \cite{elsewhere}. In this situation the mean-field equation for $\theta_\varepsilon$ remains the same as~Eq.~\eqref{eq:MF-UE}, but the distribution function $\mathcal{P}_\alpha(\textsf{a},t)$ must be found taking Fermi-liquid renormalizations into account.

The enhancement of magnetic disorder due to mesoscopic fluctuations is not restricted to classical magnetic impurities. It is known \cite{Kettemann2006,Micklitz2006,Kettemann2007,Micklitz2007} that the Kondo effect in the disordered electron systems is also modified by mesoscopic fluctuations of the local DOS. Therefore, the theory for the interplay of  the Kondo effect and superconductivity developed in Refs. \cite{MHZ1971,Matsuura1977,Bickers1987} needs to be modified for disordered
films \cite{elsewhere}.

The dependence of $T_c$ on the film conductance can be caused by a variety of reasons, among which are the dependence of the DOS at the Fermi energy on disorder, renormalization of the Cooper channel attraction in ballistic and diffusive regimes, Berezinskii-Kosterlitz-Thouless transition etc.\ \cite{Fin, BK} The sensitivity of the spin-flip rate on the conductance is a new mechanism providing a nontrivial dependence of $T_c$ on $g$.

\section{Conclusions\label{Sec6}}

To summarize, we reconsidered the problem of rare classical magnetic impurities with the Poissonian statistics in a dirty superconducting film. We took into account renormalization of the multiple spin-flip scattering due to mesoscopic fluctuations of the local DOS in a potential disorder. This effect results in the log-normal distribution of the effective magnetic impurity strength rendering the energies of quasiparticle bound states position dependent (see Fig.\ \ref{figure-levels}).
In the superconducting state, this results in the smearing of the hard gap (obtained in the absence of spin-flip renormalization) and emergence of a non-zero DOS for all energies already at the mean-field level. Depending on the bare magnetic impurity strength, the superconducting transition temperature is suppressed by the spin-flip scattering slower or faster than in the absence of renormalization. Finally, we mention that our results can be extended to the model with an arbitrary distribution of magnetic impurities, the vicinity of a superconductor-insulator transition, the case with Coulomb repulsion in addition to attraction, the presence of Zeeman splitting, etc.\ \cite{elsewhere}.

\begin{acknowledgements}
We thank M. V. Feigel'man, Ya.~V. Fominov, and A. D. Mirlin for useful discussions. The research was supported by Skoltech NGP Program (Skoltech--MIT joint project).
\end{acknowledgements}

\appendix

\section{Renormalization of the action $\mathcal S_\text{mag}$ \label{App1}}

In this Appendix we present details of the one-loop renormalization of the magnetic-impurity part of the action. 

To renormalize $\mathcal S_\text{mag}$ we write $Q=\Lambda(1+W+\dots)$, where $W$ obeys two linear constraints: $\Lambda W+W \Lambda=0$ and $W=-\overline W$, and the convergency condition $W=-W^\dagger$. The matrix $\Lambda$ is assumed to be self-dual: $\Lambda=\overline\Lambda$.  Then the quadratic part of the action reads:
\begin{equation}
\label{S0fast}
  \mathcal S_D^{(2)}[W]
  =
  -
  \frac{\pi\nu D}8 \int d \mathbf r \tr (\nabla W)^2 .
\end{equation}
The quadratic part of the action determines the following contraction rules:
\begin{subequations}
\label{contr}
\begin{gather}
  \partial_t \corr{\tr AW \tr BW}
  =
  \tr
  \left[
    AB - A\Lambda B\Lambda
  + A\overline B - A\Lambda \overline B\Lambda
  \right] ,
\\
  \partial_t \corr{\tr AWBW}
  =
    \tr A \tr B -  \tr A\Lambda \tr B\Lambda
  \notag 
\\
\hspace{26mm}
  -
  \tr
  \left[
    A\overline B - A\Lambda \overline B\Lambda
  \right] ,
\end{gather}
\end{subequations}
where $t=[2/(\pi g)] \ln (L/l)$ 
with $L$ denoting the infrared length scale.

Next we write the matrix $Q$ as $Q=U^{-1}\Lambda(1+W+\dots)U$ where the slow field $U$ obeys the condition
 $\overline U = U^{-1}$. Using contraction rules \eqref{contr}, we find
\begin{subequations}
\label{contr-Q}
\begin{gather}
  \partial_t \tr A Q B Q
  =
  \partial_t \corr{\tr U A U^{-1}\Lambda W U B U^{-1}\Lambda W}
  \notag \\
  =
  \left[
   \tr A \tr B-  \tr A Q \tr B Q
  \right]
  -
  \tr
  \left[
     A \overline{B} - A Q \overline{B} Q
  \right] ,
\\
  \partial_t \tr A Q \tr B Q
  =
  \partial_t \corr{\tr U A U^{-1}\Lambda W \tr U B U^{-1}\Lambda W}
  \notag \\
  =
  -
  \tr
  \left[
    A Q B Q - A B
  - A \overline{B} + A Q \overline{B} Q
  \right]  .
\end{gather}
\end{subequations}

The action for the slow modes after integration over fast modes $W$ can be found as
\be
\mathcal S_\text{mag} \to
- \ln \corr{e^{-\mathcal S_\text{mag}}}_W  =
  \corr{\mathcal S_\text{mag}}_W
  -\ccorr{\mathcal S_\text{mag}^2}_W/2 + \dots 
  \label{eq:Smag:ren:def:111}
\ee
Here $\corr{\dots}_W$ denotes the averaging over fast modes $W$. In what follows, in the expansion in the right-hand side of Eq.~\eqref{eq:Smag:ren:def:111} we neglect all terms except the lowest order one in the impurity concentration,  $\corr{\mathcal S_\text{mag}}_W$. The smallness of the omitted terms is controlled by the condition 
$n_s \xi^2/g \ll 1$,
where $\xi\sim\sqrt{D/T_c}$ is the superconducting coherence length in the dirty limit. As we shall see below, it is the first term in the right-hand side of Eq.~\eqref{eq:Smag:ren:def:111} that is responsible for the logarithmic renormalization of $\mathcal S_\text{mag}$.

\subsection{Operators of the second order in $Q$}

In the Born approximation (first order in $\alpha$) we need to consider the operators $T_2$ and $T_{11}$ with two $Q$ matrices involved. Their contribution to  $\mathcal S_\text{mag}$ is controlled by the coefficients $\gamma_2$ and $\gamma_{11}$ with the initial conditions following from \eqref{Smag:full}: $\gamma_2(0) = 1/4$ and $\gamma_{11}(0) = -1/8$.

Using the contraction rules (\ref{contr-Q}), we find
\be
  \partial_t \begin{pmatrix}
   \hat C T_2 \\
   \hat CT_{11}
   \end{pmatrix}
  = M_2   \begin{pmatrix}
  \hat C T_2 \\
   \hat CT_{11}
   \end{pmatrix} ,
\qquad
  M_2 =   \begin{pmatrix} 1 & -1 \\
  - 2 & 0 \end{pmatrix} .
  \label{eq:111:M2}
\ee
 The operators $T_2$ and $T_{11}$ transform into each other under the renormalization. We note that under renormalization the operators with the same or fewer number of $Q$ matrices are generated only.
\color{black}
The eigenvalues of $M_2$ are equal to $2$ and $-1$. The eigenvalue $2$ corresponds to the operator $\hat{C} {T}_2 - \hat C T_{11}/2$:
\begin{equation}
\partial_t \left (\hat C  T_2 - \frac 12 \hat C T_{11} \right ) =
2 \left (\hat C T_2 - \frac 12 \hat C T_{11} \right ).
\label{eq:T2-T11:eg1}
\end{equation}
The operator $\hat{C} {T}_2 - \hat C T_{11}/2$ is known to be a pure scaling operator beyond the lowest order perturbation theory \cite{Wegner1980,Wegner1986,Wegner1987a,Wegner1987b}.

We emphasize that the operators of the second order in $Q$ enter the magnetic part of the action,  Eq.~\eqref{Smag:full}, precisely in combination $\hat{C} {T}_2 - \hat C T_{11}/2$. \color{black} This implies that
\be
\label{alpha2}
  \gamma_{2}(t) = \frac{1}{4} e^{2 t}, \qquad  \gamma_{11}(t) = -\frac{1}{8} e^{2 t} .
\ee

\begin{widetext}

\subsection{Operators of the fourth order in $Q$}

The next nontrivial order in $\alpha$ involves operators which are of the fourth order in $Q$.
Their flow is described by the system
\begin{subequations}
\label{ren4}
\begin{align}
  \partial_t \hat CT_4
  & =
  6 (\hat CT_4)
- 4 (\hat CT_{31})
- 2  (\hat CT_{22})
- 4  \alpha \hat CT_{2} ,
\\
  \partial_t \hat CT_{31}
  & =
  - 6 (\hat CT_4)
  +
  3 (\hat CT_{31})
  -  3  (\hat CT_{211})
  - 6 \alpha \hat CT_{2}
  -  3 \alpha \hat CT_{11} ,
\\
  \partial_t \hat CT_{22}
  & =
  - 8
  (\hat CT_{4})
  +
  2 (\hat CT_{22})
  -
 2 (\hat CT_{211}) ,
\\
  \partial_t \hat CT_{211}
  & =
 - 8 (\hat CT_{31})
  -
  2  (\hat CT_{22})
  +
  (\hat CT_{211})
  -
   (\hat CT_{1111})
  - 8 \alpha \hat CT_{11} ,
\\
  \partial_t \hat CT_{1111}
  & =
 - 12  (\hat CT_{211}) .
\end{align}
\end{subequations}
 The operators of the forth order in $Q$ are mixed under the renormalization. In addition,
the operators of the second order in $Q$ are generated.
\color{black}
The system of equations \eqref{ren4} can be cast in the matrix form
\begin{equation}
\partial_t
\begin{pmatrix}
  \hat CT_4 \\
  \hat CT_{31} \\
  \hat CT_{2,2}\\
  \hat CT_{211} \\
  \hat CT_{1111} \\
  \hat CT_{2} \\
  \hat CT_{11}
\end{pmatrix} = M_4 \begin{pmatrix}
  \hat CT_4 \\
   \hat CT_{31} \\
   \hat CT_{2,2}\\
   \hat CT_{211} \\
   \hat CT_{1111} \\
   \hat CT_{2} \\
   \hat CT_{11}
\end{pmatrix} ,
\qquad
M_4 = \begin{pmatrix}
6 & -4  & -2  & 0 & 0& -4\alpha & 0 \\
-6  & 3 & 0 & -3  & 0 & -6\alpha & -3\alpha \\
-8  & 0 & 2 & -2 & 0 & 0 & 0 \\
0 & -8  & -2  & 1 & -1 & 0 & -8 \alpha \\
0 & 0& 0& -12 & 0& 0& 0 \\
0 & 0& 0& 0& 0 &  1 & -1 \\
0 & 0& 0& 0& 0 & - 2 & 0
\end{pmatrix} .
\end{equation}
Here we used Eq.~\eqref{eq:111:M2}. We emphasize that the matrix $M_4$ is the upper triangular block matrix. This reflects the fact that under renormalization the operators with the same or fewer number of $Q$ matrices are generated only.
\color{black}
The matrix $M_4$ has the following eigenvalues: 12, 5, 2, 2, $-1$, $-1$, and $-6$. The largest eigenvalue $12$ corresponds to the operator $\hat CT_4 -(2/3) \hat CT_{31} -(1/4) \hat CT_{22}+(1/4) \hat CT_{211} -(1/48) \hat CT_{1111}$.
It is known that this operator is the pure scaling operator from arguments based on the group representation theory \cite{Wegner1980,Wegner1986,Wegner1987a,Wegner1987b}.
 It is worth emphasizing that the operators of the forth order in $Q$ enter the magnetic part of the action,  Eq.~\eqref{Smag:full}, precisely in the combination $\hat CT_4 -(2/3) \hat CT_{31} -(1/4) \hat CT_{22}+(1/4) \hat CT_{211} -(1/48) \hat CT_{1111}$.
This implies that the coefficients in the action \eqref{Smag-ser} are simply
\be
\label{alpha24}
\gamma_{4}(t)
  = \frac{1}{8}
  e^{12 t}, \quad  \gamma_{31}(t)
  = -\frac{1}{12}
  e^{12 t} ,\quad \gamma_{22}(t)
  = -\frac{1}{32}
  e^{12 t} ,\quad \gamma_{211}(t)
  = \frac{1}{32}
  e^{12 t} ,\quad \gamma_{1111}(t)
  = - \frac{1}{384}
  e^{12 t} .
\ee

\subsection{Renormalization of operators of arbitrary order in $Q$}

In general, one can derive the following set of renormalization group equations:
\be
  \partial_t \hat C T_n
  =
  \frac{n(n-1)}{2} \hat C T_n
  -
  \frac{n}{2} \sum_{k=1}^{n-1}
  \left(
    \hat CT_{k,n-k}
  - (-\alpha)^{\min(k,n-k)} \hat CT_{|n-2k|}
  \right) ,
\ee
\begin{multline}
  \partial_t \hat CT_{m,n}
  =
-  2   mn
  \left(
    \hat CT_{m+n}
  - (-\alpha)^{\min(m,n)} \hat CT_{|m-n|}
  \right)
  +
  \frac{m(m-1)+n(n-1)}{2} \hat CT_{m,n}  -
    \Biggl[
  \frac{m}{2} \sum_{k=1}^{m-1}
  \Bigl (
    \hat CT_{k,m-k,n} \\
  - (-\alpha)^{\min(k,m-k)} \hat CT_{|m-2k|,n}
  \Bigr )
  +
  \frac{n}{2} \sum_{l=1}^{n-1}
  \left(
    \hat CT_{m,l,n-l}
  - (-\alpha)^{\min(l,n-l)} \hat CT_{m,|n-2l|}
  \right)
  \Biggr] ,
\end{multline}
\begin{multline}
  \partial_t \hat CT_{m,n,p}
  =
-  2
  \Bigl[
    mn
  \left(
    \hat CT_{m+n,p} - (-\alpha)^{\min(m,n)} \hat CT_{|m-n|,p}
  \right)
  + mp
  \left(
    \hat CT_{m+p,n}
  - (-\alpha)^{\min(m,p)} \hat CT_{|m-p|,n}
  \right)
\\{}
  + np
  \left(
    \hat CT_{m,n+p}
  - (-\alpha)^{\min(n,p)} \hat CT_{m,|n-p|}
  \right)
  \Bigr]
  +
  \frac{m(m-1)+n(n-1)+p(p-1)}{2} (\hat CT_{m,n,p})
 \\ -
  \Biggl[
  \frac{m}{2} \sum_{k=1}^{m-1}
  \left(
    \hat CT_{k,m-k,n,p}
  - (-\alpha)^{\min(k,m-k)} \hat CT_{|m-2k|,n,p}
    \right)
  +
  \frac{n}{2} \sum_{l=1}^{n-1}
  \Bigl(
    \hat CT_{m,l,n-l,p} \\
  - (-\alpha)^{\min(l,n-l)} \hat CT_{m,|n-2l|,p}
  \Bigr)
  +
  \frac{p}{2} \sum_{s=1}^{p-1}
  \left(
    \hat CT_{m,n,s,p-s}
  - (-\alpha)^{\min(s,p-s)} \hat CT_{m,n,|p-2s|}
  \right)
  \Biggr] .
\end{multline}
and so on.  Using these equations we find for the renormalization of the action:
\begin{gather}
  \partial_t \mathcal{S}_\text{mag}
 =
  - {n}_s \int d^2\bm{r}
  \Biggl \{ \sum_{m=1}^\infty \frac{(-1)^{m-1}}{2m}\Bigl [ \frac{m(m-1)}{2}  \hat C T_m +
  \frac{m}{2} \sum_{k=1}^{m-1} (-\alpha)^{\min(k,m-k)} \hat CT_{|m-2k|}\Bigr]\notag \\
  + \frac{1}{2!} \sum_{m,n=1}^\infty \frac{(-1)^{m+n}}{4mn} (-2 mn) \Bigl [ \hat C T_{m+n} -
 (-\alpha)^{\min(m,n)} \hat C T_{|m-n|} \Bigr ]
 + \sum_{m=1}^\infty \frac{(-1)^{m}}{2m} \frac{m}{2}  \sum_{k=1}^{m-1} \hat C T_{k,m-k}
 \notag \\
 + \frac{1}{2!} \sum_{m,n=1}^\infty \frac{(-1)^{m+n}}{4mn} \Bigl [ m(m-1) \hat C T_{m,n}+
 m \sum_{k=1}^{m-1} (-\alpha)^{\min(k,m-k)} \hat CT_{|m-2k|,n} \Bigr ]\notag \\
  + \frac{1}{3!} \sum_{m,n,p=1}^\infty \frac{(-1)^{m+n+p-1}}{8mnp}(-6 mn) \Bigl [\hat C T_{m+n,p}
 -  (-\alpha)^{\min(m,n)} \hat C T_{|m-n|,p} \Bigr]
  +\dots \Biggr \}
 \notag \\
 =
  - {n}_s \int d^2\bm{r}
  \Biggl \{
    \sum_{m=1}^\infty \frac{(-1)^{m-1}}{2m} m(m-1)
      \hat C T_m
  + \frac{1}{2!} \sum_{m,n=1}^\infty \frac{(-1)^{m+n}}{4mn} (m+n)(m+n-1)
      \hat C T_{mn} + \dots
  \Biggr \}
  \label{eq:Smag:dt}
  \end{gather}
Note that all terms in Eq.~\eqref{eq:Smag:dt} which contain $\alpha$ cancel each other.

\end{widetext}

\section{The renormalized action $\mathcal S_\text{mag}$\label{App2}}

In this Appendix we present the details of derivation of Eq. \eqref{Smagn0}.

All in all, we find from Eq.~\eqref{eq:Smag:dt}  that the coefficients $\gamma_{k_1k_2\dots k_q}$, where $k_1+k_2+\dots+k_q= n$, behaves in the same way:
\be
  \gamma_{k_1k_2\dots k_q}(t)
  =
  \gamma_{k_1k_2\dots k_q}(0) e^{n(n-1)t} .
  \label{eq:multi-spectrum}
\ee

In what follows we are interested in the mean-field analysis of the renormalized action \eqref{Smag-ser} for which the singlet sector of the theory is important only. Therefore,
one can operate with $Q$ matrix which is the unit matrix in the spin space, $Q=Q_0 \sigma_0$. Then averaging over directions of the impurity magnetization $\bm{n}$ becomes trivial. We find (all indices, $m$, $n$, $\dots$ are even)
\begin{equation}
\begin{split}
  \hat C T_m
  & =
  (-\alpha)^{m/2} \tr (Q \tau_3)^m,  \\
  \hat C T_{mn}
  & =
  (-\alpha)^{(m+n)/2} \tr (Q \tau_3)^m \tr (Q \tau_3)^n,
\end{split}
\end{equation}
and so on. Then the renormalized action for magnetic impurities becomes
\begin{widetext}
\begin{multline}
 \mathcal S_\text{mag}
 =
  - {n}_s \int d^2\bm{r}
  \Biggl[
  - \sum_{k=1}^\infty \frac{(-\alpha)^{k}}{2^2k}
  e^{2k(2k-1)t}
  \tr (Q \tau_3)^{2k}
  +
  \frac{1}{2!} \sum_{k,l=1}^\infty \frac{(-\alpha)^{k+l}}{4^2kl}
  e^{(2k+2l)(2k+2l-1)t}
  \tr (Q \tau_3)^{2k} \tr (Q \tau_3)^{2l}
\\{}
  -
  \frac{1}{3!} \sum_{k,l,m=1}^\infty \frac{(-\alpha)^{k+l+m}}{8^2klm}
  e^{(2k+2l+2m)(2k+2l+2m-1)t}
  \tr (Q \tau_3)^{2k} \tr (Q \tau_3)^{2l} \tr (Q \tau_3)^{2m}
  + \dots
  \Biggr] .
\end{multline}
Decoupling the Gaussian part with an auxiliary integral over $\lambda$ we obtain
\begin{multline}
\mathcal  S_\text{mag}
 =
  - \overline{n}_s \int d^2\bm{r}
  \int \frac{d\lambda}{\sqrt{4\pi t}} e^{-(\lambda+t)^2/4t}
  \Biggl[
  - \sum_{k=1}^\infty \frac{(-\alpha)^{k}}{2^2k}
  e^{2k\lambda}
  \tr (Q \tau_3)^{2k}  +
 \sum_{k,l=1}^\infty \frac{(-\alpha)^{k+l}}{2!4^2kl}
  e^{(2k+2l)\lambda}
  \tr (Q \tau_3)^{2k} \tr (Q \tau_3)^{2l}
\\{}
  -
  \sum_{k,l,m=1}^\infty \frac{(-\alpha)^{k+l+m}}{3!8^2klm}
  e^{(2k+2l+2m)\lambda}
  \tr (Q \tau_3)^{2k} \tr (Q \tau_3)^{2l} \tr (Q \tau_3)^{2m}
  + \dots
  \Biggr] .
\end{multline}
Now all summations become trivial:
\begin{gather}
\mathcal  S_\text{mag}
  =
  - {n}_s \int d^2\bm{r}
  \int \frac{d\lambda}{\sqrt{4\pi t}} e^{-(\lambda+t)^2/4t}
  \Biggl[
  X
  +
  \frac{X^2}{2!}
  +
  \frac{X^3}{3!}
  + \dots
  \Biggr]
  =
  - {n}_s \int d^2\bm{r}
  \int \frac{d\lambda}{\sqrt{4\pi t}} e^{-(\lambda+t)^2/4t}
  (e^X-1) ,
\end{gather}
\end{widetext}
where
\begin{gather}
  X
  =-
  \tr \sum_{k=1}^\infty \frac{(-\alpha)^{k}}{4k}
  e^{2k\lambda}
  (Q \tau_3)^{2k}
 \notag 
\\ =
  \frac14 \tr \ln\bigl [1+ \alpha e^{2\lambda} (Q \tau_3)^{2}\bigr ]  .
   \label{eq-Smag-S25}
\end{gather}
Finally, we find
\begin{align}
\mathcal  S_\text{mag}
  = & 
  - {n}_s \int d^2\bm{r}
  \int \frac{d\lambda}{\sqrt{4\pi t}}
  \exp\left\{-\frac{(\lambda+t)^2}{4t}
  \right\}
  \notag \\
  & \times 
  \left[
  \exp\left\{
    \frac14 \tr \ln[1+ \alpha e^{2\lambda} (Q \tau_3)^{2}]
  \right\}
  - 1
  \right] ,
\end{align}
where $\tr$ still includes summation over the spin space. 
This equation is equivalent to Eq. \eqref{Smagn0}.

\section{The spin-flip rate, the transition temperature, and the density of states\label{App3}}

In this Appendix we present the details of derivation of results for the spin-flip rate, the transition temperature, and the density of states.

\subsection{The spin-flip rate near the transition temperature\label{App3:1}}

According to Eq. \eqref{eq:SFR}, contrary to the usual case of magnetic disorder \cite{AG2}, the spin-flip rate in the presence of mesoscopic fluctuations acquires a weak logarithmic (in 2D) dependence on energy through the function $t=t(\varepsilon)$:
\begin{align}
\frac{1}{\tau_s} & =
\frac{2 n_s}{\pi \nu}\int \limits_{-\infty}^\infty \frac{du}{\sqrt\pi} e^{-u^2-2 i \mu u} \int \limits_{0}^\infty \frac{d\lambda}{\sqrt{\pi}} \frac{\cos(2 u \lambda)}{\cosh (\beta\lambda)+1} 
\notag \\
& =\frac{2 n_s}{\pi \nu} \frac{4}{\beta^2}\int \limits_{-\infty}^\infty du \, e^{-u^2-2i u \mu} \frac{u}{\sinh\frac{2\pi u}{\beta}} .
\label{eq:SFR-S}
\end{align}
Here we introduced $\mu = (2t-\ln\alpha)/(4\sqrt{t})$ and $\beta=4\sqrt{t}$. Expanding the denominator in the last integral in the right hand side  of Eq.~\eqref{eq:SFR-S} in powers of $\exp(-2\pi |u|/\beta)$ and, then, performing integration over $u$, we find
\begin{equation}
\frac{1}{\tau_s} = - \frac{2 n_s}{\pi \nu} \frac{2\sqrt{\pi}}{\beta^2} \partial_\mu \im \sum_{k=0}^\infty f\bigl (i\mu+\pi(2k+1)/\beta\bigr )
.
\label{eq:tau-s0-S}
\end{equation}
Here we introduce the function $f(z)=\exp(z^2)[1-\erf(z)]$. At $\beta/\pi \ll 1$ we can use the expansion of the function $f(z)$ in
series in $1/z$:
\begin{equation}
f(z) =\sum_{l=1}^\infty \frac{(-1)^{l-1} \Gamma(l-1/2)}{\pi z^{2l-1}} .
\label{eq:asym:f}
\end{equation}
Performing summation over $k$ in \eqref{eq:tau-s0-S}, we obtain
\begin{gather}
\frac{1}{\tau_s} = \frac{n_s}{\pi \nu \beta}\sum_{l=0}^\infty \frac{1}{4^l l!} \partial_\mu^{2l+1} \tan \left (\frac{\beta \mu}{2}\right ) \notag \\
=
\frac{2 n_s}{\pi \nu}e^{4t (\alpha \partial_\alpha)^2} \frac{\alpha e^{-2t}}{(1+\alpha e^{-2t})^2}
.
\label{eq:tau-s-S}
\end{gather}
\color{black}

In the limiting cases Eq.\ \eqref{eq:tau-s-S} reduces to
\begin{equation}
\frac{1}{\tau_s} =  \frac{2 n_s}{\pi \nu} \begin{cases}
 \alpha e^{2t}, &  \alpha \to 0 ,\\
 e^{6t}/\alpha, &  \alpha \to \infty .
\end{cases}
\end{equation}
Expanding the right hand side of Eq.\ \eqref{eq:tau-s-S} to the first order in $t$, we find
\begin{equation}
\frac{1}{\tau_s}= \frac{1}{\tau_{s0}}\left [ 1 + 2 t \frac{1-8\alpha+3\alpha^2}{(1+\alpha)^2}+ O(t^2)\right ] ,
\end{equation}
where $1/\tau_{s0} = 2 \alpha n_s/[\pi \nu (1+\alpha)^2]$.

For $\beta/\pi \gg 1$, the sum $k$ in Eq.\ \eqref{eq:tau-s0-S} reduces to the integral. Then, we find
\begin{gather}
\frac{1}{\tau_s} = \frac{2 n_s}{\pi \nu}\frac{1}{\sqrt{\pi} \beta}\re f\bigl (i\mu+\pi/\beta\bigr ) .
\end{gather}
Next, for $\mu\beta \gg 1$, which holds for $t\gg 1$, we obtain
\begin{equation}
\frac{1}{\tau_s} = \frac{2 n_s}{\pi \nu}\frac{1}{\sqrt{\pi} \beta} e^{-\mu^2} = \frac{1}{\tau_{s0}} \frac{(1+\alpha)^2}{\alpha} \mathcal{P}_\alpha(1,t) .
\end{equation}

\subsection{The transition temperature\label{App3:2}}

Knowledge of the effective spin-flip rate allows one to compute the dependence of the superconducting transition temperature on the spin-flip rate, $1/\tau_{s0}$, and potential disorder. Using Eqs. \eqref{eq:Supp:SCE} and \eqref{eq:Delta:lin}, we find the following equation for the transition temperature:
\begin{equation}
\ln \frac{T_{c0}}{T_c} = \sum_{n=0}^\infty \left [ \frac{1}{n+1/2+1/(2\pi T_c \tau_s(\varepsilon))} - \frac{1}{n+1/2} \right ] ,
\label{eq:Supp:l1}
\end{equation}
where $\varepsilon=2\pi T_c(n+1/2)$. Performing formal expansion in the right hand side of Eq. \eqref{eq:Supp:l1} with respect of the difference  $\tau_s^{-1}(\varepsilon)-\tau_s^{-1}(T_c)$, we obtain
\begin{equation}
\ln \frac{T_{c0}}{T_c} = \psi\left (\frac12+\frac{1}{2\pi T_c \tau_s(T_c)}\right ) - \psi\left (\frac12\right )+ X_{\rm rest},
\end{equation}
where 
\begin{equation}
X_{\rm rest}=
\sum_{n=0}^\infty\sum_{k=1}^\infty \frac{(-1)^{k+1} \left (\tau^{-1}_s(\varepsilon)-\tau_s^{-1}(T_c) \right )^k }{(2\pi T_c)^k [n+1/2+1/(2\pi T_c \tau_s(T_c))]^{k+1}}
.
\label{eq:Supp:l2}
\end{equation}
Since the effective spin-flip rate depends on the Matsubara energy $\varepsilon$ via $t(\varepsilon) = t(T_c) - \frac{1}{\pi g} \ln [\pi (2n+1)]$, we can represent  $\tau_s^{-1}(\varepsilon)$ as follows
\begin{equation}
\frac{1}{\tau_s(\varepsilon)} = \frac{1}{\tau_{s}(T_c)} + \sum_{l=1}^\infty \frac{(-1)^l}{l! (\pi g)^l}   \ln^l \bigl [ \pi (2n+1) \bigr ] \frac{\partial^l }{\partial t^l} \frac{1}{\tau_{s}} \Biggl |_{t=t(T_c)} .
\label{eq:Supp:l3}
\end{equation}

In the case $1/(2\pi T_c\tau_s) \ll 1$, the sum in Eq. \eqref{eq:Supp:l2} is dominated by the term with $k=1$. The condition $g\gg 1$ allows us to consider in Eq. \eqref{eq:Supp:l2} the term with $l=1$ only. Therefore, we find
\begin{equation}
X_{\rm rest} =  \frac{c}{g T_c\tau_s(T_c)} \frac{\partial \ln \tau_s}{\partial t} \Biggl |_{t=t(T_c)} , \qquad  \frac{1}{2\pi T_c\tau_s} \ll 1  ,
\label{eq:Supp:l3a}
\end{equation}
where numerical constant 
\begin{equation}
c = \frac{1}{2\pi^2} \sum_{n=0}^\infty \frac{\ln[\pi(2n+1)]}{(n+1/2)^2} \approx 0.4  .
\end{equation}
Using Eq. \eqref{eq:Supp:l1}, we find that the suppression of $T_c$ for the case $1/(2\pi T_c\tau_s) \ll 1$ is given as
\begin{equation}
\frac{T_c-T_{c0}}{T_{c0}} = \left (1+\frac{4 c}{\pi g} \right ) \frac{\pi}{4 T_{c0} \tau_s(T_{c0})} .
\end{equation}
As one can see,  the correction to the expression for $T_c$ due to the dependence of the effective spin-flip rate on the Matsubara energy is negligible provided the conductance is large enough, $g \gg 1$.

In the opposite case, $1/(2\pi T_c\tau_s) \gg 1$, we can integrate over $n$ in Eq. \eqref{eq:Supp:l2} instead of summation and find
 \begin{gather}
X_{\rm rest}=
\sum_{k=1}^\infty \frac{(-1)^{k+1}}{k} \left (\sum_{l=1}^\infty \frac{\tau_s \ln^l \bigl [2\pi T_c\tau_s\bigr ] }{l! (\pi g)^l}   \frac{\partial^l }{\partial t^l} \frac{1}{\tau_{s}} \Biggl |_{t=t(T_c)} \right )^{k} \notag \\
=
\ln \frac{\tau_s(t(T_c))}{\tau_s\bigl (t(T_c)+\frac{1}{\pi g} \ln \frac{1}{2\pi T_c\tau_s(t(T_c))}\bigr )}
.
\label{eq:Supp:l2}
\end{gather}
Provided the following condition 
\begin{equation}
\frac{1}{\pi g} \ln \frac{1}{2\pi T_c\tau_s} \ll 1
\label{eq:Supp:l4}
\end{equation}
holds, we can neglect the term $X_{\rm rest}$ in the right hand side of Eq. \eqref{eq:Supp:l1} in comparison with the $\ln 1/(2\pi T_c\tau_s)$ which appears due to the di-gamma function. 

All in all, the correction $X_{\rm rest}$ to the mean-field equation \eqref{eq:Supp:l1} for $T_c$ which stems from the energy dependence of the effective spin-flip rate can be neglected if the following inequality holds:
\begin{equation}
\frac{1}{g} \max\left \{1, \ln \frac{1}{2\pi T_c\tau_s}\right \} \ll 1 .
\label{eq:Supp:l45}
\end{equation}
Since our theory is valid for $t(T_c) = [1/(\pi g)] \ln [{1}{(2\pi T_c\tau)}] \ll 1$ and $1/\tau_s\ll 1/\tau$, the condition  \eqref{eq:Supp:l45} is always satisfied.

\subsection{The density of states\label{App3:3}}

The average DOS can be extracted from $\corr{Q_{\varepsilon\varepsilon}}$ analytically continued to the real energies $E$: $i\varepsilon \to E+i 0^+$. The mean-field equation \eqref{eq:MF-UE} can be written as
\begin{equation}
\varepsilon \sin \theta_\varepsilon - \Delta \cos \theta_\varepsilon + \frac{n_s}{\pi \nu} \mathcal{F}\left (\theta_\varepsilon, \frac{2t-\ln\alpha}{4\sqrt{t}},4\sqrt{t}\right ) = 0,
\label{eq:MFUE-11}
\end{equation}
where
\begin{equation}
\mathcal{F}(\theta,\mu,\beta) = \int \limits_{-\infty}^\infty \frac{d\lambda}{2\sqrt{\pi}} e^{-(\lambda+\mu)^2}\frac{\sin2\theta}{\cosh (\beta\lambda)+\cos2\theta}
.
\end{equation}
It is convenient to parametrize the spectral angle as $\theta_{\varepsilon} = \pi/2+ i\psi$ such that the density of states becomes:
 \begin{equation}
 \rho(E)= 2\nu \lim_{i \varepsilon \to E+i0^+} \im \sinh \psi .
\end{equation}
For arbitrary values of $t$ and $\alpha$, Eq.~\eqref{eq:MFUE-11} is a complicated integral equation which can be solved numerically. Below, we demonstrate how its solution and, consequently, the density of states, can be found analytically at $t\ll 1$.

At first, we rewrite the function $\mathcal{F}(\theta,\mu,\beta)$ as follows
\begin{gather}
\mathcal{F}(\theta,\mu,\beta) = \int \limits_{-\infty}^\infty \frac{du}{\sqrt\pi} e^{-u^2-2 i \mu u} \int \limits_{0}^\infty \frac{d\lambda}{\sqrt{\pi}} \frac{\sin2\theta \cos(2 u \lambda)}{\cosh (\beta\lambda)+\cos2\theta} \notag \\
=
\frac{1}{\beta} \int \limits_{-\infty}^\infty du \ \frac{\sinh \frac{4\theta u}{\beta}}{\sinh\frac{2\pi u}{\beta}} \ e^{-u^2-2i u \mu} .
\label{eq:F:def-23}
\end{gather}
Here we used the relation 3.514.2 from the book \cite{GR}. Expanding the denominator in the last integral in the right hand side  of Eq.~\eqref{eq:F:def-23} in powers of $\exp(-2\pi |u|/\beta)$ and, then, performing integration over $u$, we find
\begin{gather}
\mathcal{F}(\pi/2+i\psi,\mu,\beta)= \frac{\sqrt\pi}{2\beta} \sum_{\sigma=\pm}\Biggl \{ \sum_{k=0}^\infty
f\left (i \sigma \mu+\frac{2(\pi k - i\psi)}{\beta}\right ) 
\notag \\
- \sum_{k=1}^\infty f\left (i\sigma \mu+\frac{2(\pi k+ i \psi)}{\beta}\right )
\Biggr \}
,
\label{eq:F:def:456}
\end{gather}
where $f(z) = \exp(z^2)[1-\erf(z)]$. Since we are interested in $\beta/\pi\lesssim 1$ we can use expansion of the function $f(z)$ in powers of $1/z$ (see Eq.~\eqref{eq:asym:f}). Then performing summation over $k$ in the right hand side of Eq.~\eqref{eq:F:def:456}, we find
\begin{gather}
\mathcal{F}(\pi/2+i\psi,\mu,\beta) =
\frac{1}{2\beta}\Bigl [ H(\mu - 2 \psi/\beta)+H(-\mu - 2 \psi/\beta)\Bigr ] \notag \\
- \frac{1}{2} e^{\frac{1}{4} \partial_\mu^2}\frac{i \sinh (2\psi)}{\cosh(\beta \mu)-\cosh(2\psi) }
,
\label{eq:mathF-S}
\end{gather}
where
\begin{equation}
H(z) = \sqrt\pi e^{-z^2} \bigl [ 1- i \erfi(z) \bigr ] +  i\ e^{\frac{1}{4} \partial_z^2} \, z^{-1}  .
\end{equation}
While deriving Eq.~\eqref{eq:mathF-S} we used the following relation for the Euler di-gamma functions:
\begin{equation}
\psi(1+z)-\psi(1-z)=\frac{1}{z} -\frac{\pi}{\tan\pi z}   .
\end{equation}
We note that both the real and imaginary parts of the function $H(z)$ are exponentially small at
 $z\gg 1$.

Using the result \eqref{eq:mathF-S} and making transformation $\varepsilon \to - i E$, we obtain the following form of the mean-field equation \eqref{eq:MF-UE}:
\begin{gather}
e^{4t (\alpha \partial_\alpha)^2}F_E(\psi,\alpha e^{-2t})  = \frac{i n_s}{8\pi \sqrt{t}\ \nu \Delta} \biggl [ H\Bigl(\frac{2t-\ln \alpha-2\psi}{4\sqrt{t}} \Bigr)
\notag 
\\ {}
+H\Bigl(\frac{\ln \alpha-2t-2\psi}{4\sqrt{t}} \Bigr)\biggr ] ,
\label{eq:MFE:tt}
\end{gather}
where the function (cf. Eq. \eqref{eq:MF-UE-t0})
\begin{equation}
F_E(\psi,\alpha)= \sinh \psi -\frac{E}{\Delta} \cosh\psi - \frac{[\alpha n_s/(\pi \nu \Delta)]\sinh 2\psi}{1+\alpha^2 -2\alpha \cosh 2 \psi}
\end{equation}
is defined in such a way that the mean-field equation at $t=0$ is given as
\begin{equation}
F_E(\psi,\alpha)=0 .
\label{eq:MFE:t0}
\end{equation}

In what follows, we focus at the case
\begin{equation}
\frac{1}{\tau_{s0}\Delta}< \frac{(1-\alpha)}{(1+\alpha)}^2 ,
\end{equation}
in which the density of states has a finite gap $E_{g0}$ at $t=0$ \cite{FS2016}. In this case, Eq.~ \eqref{eq:MFE:t0} has a real solution $\psi$ for $|E|<E_{g0}$. The energy $E_{g0}$ and the corresponding value $\psi_{g0}$ are determined from the following equations:
\begin{equation}
F_{E_{g0}}(\psi_{g0},\alpha)=0 , \qquad \partial_{\psi_{g0}} F_{E_{g0}}(\psi_{g0},\alpha)=0 .
\end{equation}

Since for $t\ll 1$ the arguments of the functions $H$ in Eq. \eqref{eq:MFE:tt} are large we can use the asymptotic expression for $H(z)$ at $z\gg 1$. In this way, we find
\begin{equation}
\begin{split}
\widetilde{F}_E(\psi,\alpha,t)  = \frac{i n_s}{2 \nu \Delta}\sum\limits_{\sigma=\pm} e^{2\psi \sigma}\mathcal{P}_{\alpha}(e^{2\psi \sigma},t)  , \\  \widetilde{F}_E(\psi,\alpha e^{-2t}) \equiv e^{4t (\alpha \partial_\alpha)^2}F_E(\psi,\alpha e^{-2t})  .
\end{split}
\label{eq:MFE:tt-2-1}
\end{equation}
We note that for $t\ll 1$, we can write
\begin{gather}
\widetilde{F}_E(\psi,\alpha,t) \approx \bigl [ 1 -2 t (\alpha \partial_\alpha) + 4t (\alpha \partial_\alpha)^2 \Bigr ] F_E(\psi,\alpha) .
\end{gather}

\subsubsection{The density of states near the band gap $E_{g0}$}

The solution of Eq.~\eqref{eq:MFE:tt} for $t\ll 1$ depends on the energy interval we are interested in. We start from the energies close to the bare gap edge $E_{g0}$. The function $\widetilde{F}_E(\psi,\alpha,t)$ has similar behaviour as the function $F_E(\psi,\alpha)$. Although at nonzero $t$ the density of states is finite at some energy, it is convenient to define the 
characteristic energy $E_g$ and corresponding angle $\psi_g$ which are the solutions of the following set of equations:
 \begin{equation}
\widetilde{F}_{E_{g}}(\psi_{g},\alpha,t)=0 , \qquad \partial_{\psi_{g}} \widetilde{F}_{E_{g}}(\psi_{g},\alpha,t)=0 .
\end{equation}
For $t\ll 1$ we find that the difference between the characteristic energy $E_g$ and the bare gap $E_{g0}$ is given as
\begin{gather} 
E_{g0}-E_{g} = \frac{2t \Delta}{\cosh\psi_{g0}} \bigl [  (\alpha \partial_\alpha) -2  (\alpha \partial_\alpha)^2 \Bigr ] F_{E_{g0}}(\psi_{g0},\alpha) .
\end{gather}

In the Abrikosov-Gor'kov regime, $\alpha\ll \eta^{2/3}\ll 1$, where $\eta \equiv 1/(\tau_{s0\Delta})$, the above expression for the shift of the bare gap acquires the following simple form:
\begin{equation}
\frac{E_{g0}-E_{g}}{E_{g0}} = 2t \eta^{2/3} .
\label{eq:AG:shift}
\end{equation}
Here we took into account that $\cosh\psi_{g0} = 1/\eta^{1/3}$ and $E_{g0} = \Delta (1-\eta^{2/3})^{3/2}$.

Now we can find the dependence of the density of states on energy near $E_g$. Expanding the left hand side of Eq. \eqref{eq:MFE:tt-2-1} in $\epsilon = (E-E_g)/\Delta$ and $\psi-\psi_g$, we find the following result for the density of states:
\begin{gather}
\frac{\rho(E)}{2\nu} = \cosh \psi_g \sqrt{\frac{2\cosh\psi_g}{|\partial_{\psi_g}^2\widetilde{F}_{E_g}(\psi_g,\alpha,t)|}}
\, {\textrm{Re}\,} \sqrt{\epsilon+i \epsilon_*}, \notag \\ |\epsilon| \ll \frac{|\partial_{\psi_g}^2\widetilde{F}_{E_g}(\psi_g,\alpha,t)|}{\cosh\psi_g} .
\end{gather}
Here we introduced the energy scale
\begin{equation}
\epsilon_* = \frac{n_s}{2 \nu \Delta \cosh\psi_g}\sum\limits_{\sigma=\pm} e^{2\psi_g \sigma}\mathcal{P}_{\alpha}(e^{2\psi_g \sigma},t) .
\end{equation}

In the regime $\alpha\ll \eta^{2/3}\ll 1$ the result for the density of states for $|\epsilon|\ll \eta^{2/3}$ becomes
\begin{gather}
\frac{\rho(E)}{2\nu} = \frac{\sqrt{2}}{\eta^{2/3}\sqrt{3}}\, {\textrm{Re}\,} \sqrt{\epsilon+i \epsilon_*}\ ,
\notag\\
 \epsilon_* = \frac{\eta^{4/3}\sqrt\pi}{16 \alpha \sqrt{t}}
\left (\frac{4\alpha}{\eta^{2/3}}\right )^{1/4} \exp \left ( -\frac{1}{16 t}\ln^2\frac{\eta^{2/3}}{\alpha} \right ) .
\label{eq:our:rho:app}
\end{gather}  

Now it is instructive to compare our results for the density of states with the results of the instanton analysis \cite{FS2016,MaS}. The density of states due to instantons near the band gap $E_g$ is given as
\begin{equation}
\frac{\rho_{\rm inst}(\epsilon)}{2\nu} \approx \frac{\cosh\psi_g}{\sqrt{g}} \exp \Bigl ( -g \frac{2\cosh\psi_g}{|\partial_{\psi_g}^2\widetilde{F}_{E_g}(\psi_g,\alpha,t)|} |\epsilon| \Bigr ) .
\label{eq:inst:rho:app}
\end{equation}
As one can see there is the characteristic energy scale 
$\Gamma= {|\partial_{\psi_g}^2\widetilde{F}_{E_g}(\psi_g,\alpha,t)|}/({g \cosh\psi_g})$ in Eq. \eqref{eq:inst:rho:app}. 
Using Eq. \eqref{eq:our:rho:app}, we find
\begin{equation}
\frac{\rho(\Gamma)}{\rho_{\rm inst}(\Gamma)}
\sim
\begin{cases}
\epsilon_*/\Gamma , & \quad \epsilon_* \ll \Gamma , \\
\sqrt{\epsilon_*/\Gamma} , & \quad \epsilon_* \gg \Gamma .
\end{cases}
\end{equation}
Therefore, our contribution to the density of states dominates the instanton one near the band gap $E_g$ provided $\epsilon_* \gg \Gamma$. In the Abrikosov-Gor'kov regime,
$\alpha\ll \eta^{2/3}\ll 1$, this condition becomes
\begin{equation}
\frac{1}{\sqrt{t}}\exp\left ( -\frac{1}{16 t} \ln^2 \frac{\eta^{2/3}}{\alpha}\right ) \gg \frac{1}{g}\left ( \frac{\alpha}{\eta^{2/3}}\right )^{3/4}  .
\end{equation}
At the Fermi level our contribution to the density of states dominates the result due to instanton analysis since the
latter involves the sheet resistance $1/g$ which is parametrically smaller than \emph{spreading resistance} $t = \ln(\xi/l)/(2\pi g)$.

\begin{figure}
\centerline{\includegraphics[width=0.93\columnwidth]{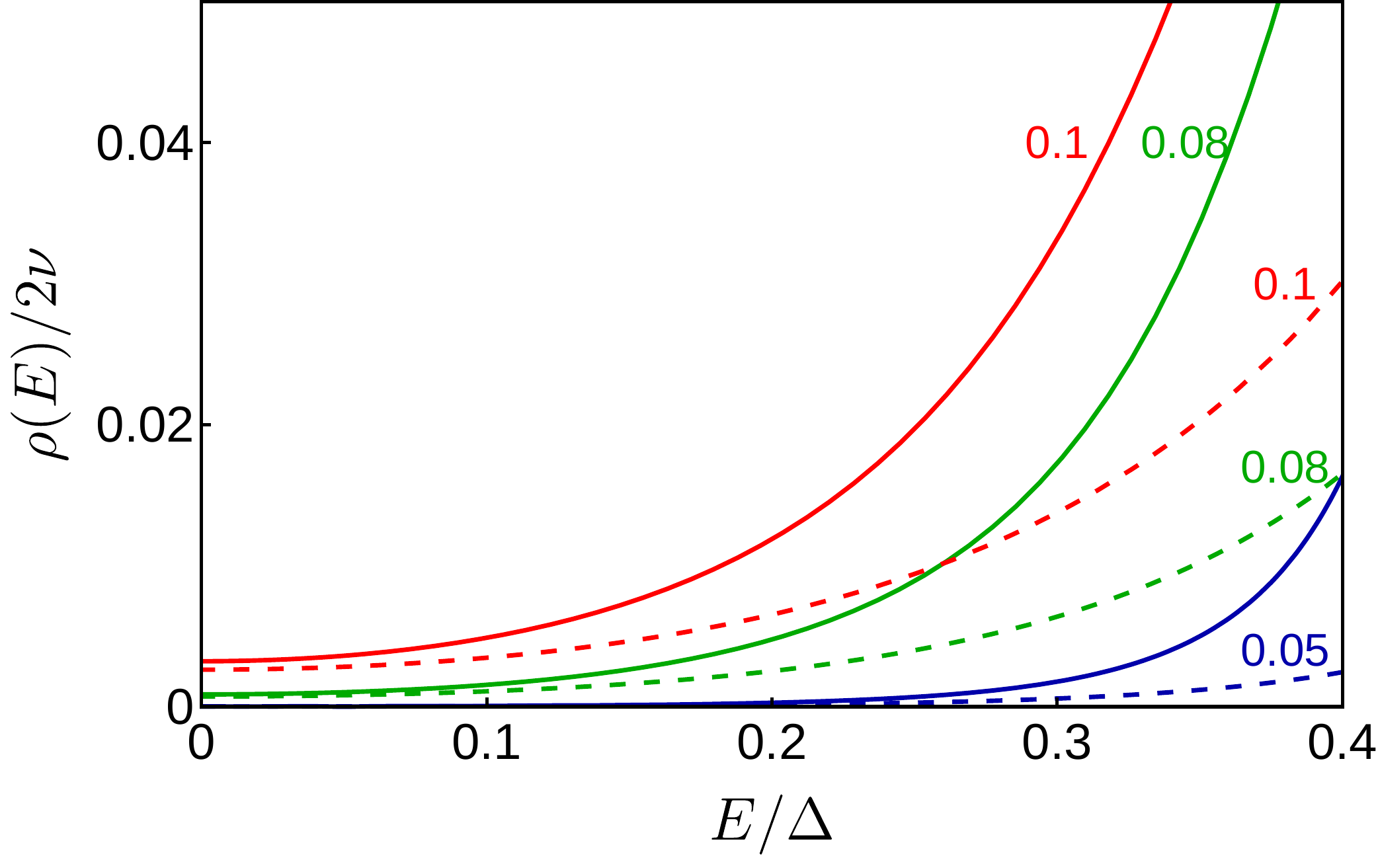}}
\caption{Energy dependence of the density of states for some values of the parameter $t$. The solid curves are obtained by numerical solution of the mean-field Eq.~\eqref{eq:MFUE-11}. The dashed curves are plotted with the help of Eqs. \eqref{eq:DOS-f-S}. We use $1/(\tau_{s0}\Delta) = 0.1$ and $\alpha=0.05$.
}
\label{figure-DOS-sup}
\end{figure}

\subsubsection{The density of states at low energies}

At energies which are much smaller than the characteristic energy, $|E|\ll E_{g}$, the equation  \eqref{eq:MFE:tt-2-1} without the right hand side has the real solutions only. We substitute  $\psi=\psi^\prime+i \psi^{\prime\prime}$ with $\psi^{\prime\prime} \ll 1$ into  Eq.~\eqref{eq:MFE:tt-2-1} and splitting into the real and imaginary parts. Then we find
\begin{equation}
\begin{split}
\widetilde{F}_E(\psi^\prime,\alpha,t) & =0, \notag \\
\partial_{\psi^\prime} \widetilde{F}_E(\psi^\prime,\alpha,t) \psi^{\prime\prime} & = \frac{n_s}{2 \nu \Delta}\sum\limits_{\sigma=\pm} e^{2\psi^\prime \sigma}\mathcal{P}_{\alpha}(e^{2\psi^\prime\sigma},t) 
.
\end{split}
\label{eq24}
\end{equation}
The density of states can be found as
\begin{gather}
\frac{\rho(E)}{2\nu}= \psi^{\prime\prime} \cosh\psi^\prime  = \frac{n_s}{2\nu\Delta}\frac{\cosh\psi^\prime}{\partial_{\psi^\prime} \widetilde{F}_E(\psi^\prime,\alpha,t)}
\notag \\
\times
\sum\limits_{\sigma=\pm} e^{2\psi^\prime \sigma}\mathcal{P}_{\alpha}(e^{2\psi^\prime\sigma},t)
.
\label{eq:DOS-f-S}
\end{gather}
We present the comparison between the density of states found from numerical solution of Eq.~\eqref{eq:MFUE-11} and analytical result \eqref{eq:DOS-f-S} in Fig. \ref{figure-DOS-sup}. To plot the curves in this figure we neglect the difference between $\psi_g$ and $\psi_{g0}$ as well as between $\widetilde{F}_E(\psi^\prime,\alpha,t)$ and $F_E(\psi^\prime,\alpha)$. 

At $E=0$, $\psi^\prime= 0$ is the solution of Eq.~\eqref{eq24}. Then from Eq.\ \eqref{eq:DOS-f-S}  we find the density of states at zero (Fermi) energy
\begin{align}
\rho(0) = {} & \frac{2 n_s}{\Delta} \left (1- \frac{1}{\tau_{s0}\Delta} \frac{ (1+\alpha)^2}{(1-\alpha)^2
} \Bigl [ 1+2t \frac{1+8\alpha+3\alpha^2}{(1-\alpha)^2}\Bigr ]\right )^{-1} 
\notag 
\\ & {}
\times \mathcal{P}_\alpha(1,t)
.
\end{align}

\section{The effect of termination of the multifractal spectrum\label{App4}}

In this Appendix we discuss how the termination of the multifractal spectrum affects our results.

The result \eqref{eq:multi-spectrum} for the coefficients $\gamma_{k_1k_2\dots k_q}$ is derived by consideration of the contributions related with $\langle \mathcal{S}_\textrm{mag}\rangle_W$. In this approximation operators $T_{k_1k_2\dots k_q}$ with given $n=k_1+\dots + k_q$ always transform under the renormalization group into linear combinations of operators  $T_{l_1l_2\dots l_q}$ with $m=l_1+\dots+l_q \leqslant n$.
Therefore, the renormalization group equations remain linear in coefficients $\gamma_{k_1k_2\dots k_q}$. In general, one needs to take into account terms which are nonlinear in $\mathcal{S}_\textrm{mag}$, e.g.\ $\langle [\mathcal{S}_\textrm{mag}]^2\rangle_W$. Then the fusion of two operators  $T_{k_1k_2\dots k_q}$ and $T_{l_1l_2\dots l_q}$ into a single operator $T_{s_1s_2\dots s_q}$ with $s_1+\dots+s_q=n+m$ is possible. This renders the renormalization group equations for $\gamma_{k_1k_2\dots k_q}$ nonlinear \cite{Foster2009}. This nonlinearity results in termination of the multifractal spectrum \cite{Mirlin2000} which implies the following modification of Eq.~\eqref{eq:multi-spectrum}:
\begin{gather}
  \gamma_{k_1k_2\dots k_q}(t)
  =
  \gamma_{k_1k_2\dots k_q}(0) e^{\,\textsf{y}_n t}, \notag \\
 \textsf{y}_n = \begin{cases}
  n(n-1) , \quad & 1<n<n_c ,\\
  -n_c^2 + (2n_c-1) n , \quad & n_c\leqslant n .
\end{cases}
  \label{eq:multi-spectrum-t}
\end{gather}
Here $n_c=\sqrt{2/t_0} \gg 1$ and $t_0 = 2/(\pi g) \ll 1$ denotes the bare resistance. The function $\textsf{y}_n$ obeys the following symmetry property: $\textsf{y}_{1-n}=\textsf{y}_n$ \cite{Gruzberg2013}.  Let us now define the function $\mathcal{G}(\lambda)$ as
\begin{equation}
\int \limits_{-\infty}^{\infty} d\lambda \, e^{n \lambda} \, \mathcal{G}(\lambda) = e^{\,\textsf{y}_{n} t} .
\end{equation}
Then we find
 \begin{gather}
 \mathcal{S}_\textrm{mag} =
  - {n}_s \int d^2\bm{r}
  \int \limits_{-\infty}^{\infty} d\lambda \, \mathcal{G}(\lambda)
  \bigl (e^X-1 \bigr ) ,
\end{gather}
where $X$ is given by Eq.\ \eqref{eq-Smag-S25}.

At $t\ll 1$ and $t/t_0\gg 1$, the function $\mathcal{G}(\lambda)$ can be written as
\begin{equation}
\mathcal{G}(\lambda) = \frac{1}{\sqrt{4\pi t}} \exp\left [-\frac{(\lambda+t)^2}{4t} \right ] \theta(\lambda_c-|\lambda|),
\end{equation}
 where $\theta(z)$ denotes the Heaviside step function and $\lambda_c = t (2 \sqrt{2/t_0} - 1) \approx 2 t\sqrt{2/t_0}$. This form of  the function $\mathcal{G}(\lambda)$ implies that the integration over $\textsf{a}$ in Eq.\ \eqref{eq:MF-UE} is restricted to the range $\textsf{a}_- < \textsf{a} < \textsf{a}_+$, where $\textsf{a}_\pm = \alpha \exp (\pm 2\lambda_c)$.
Since for the existence of a finite density of states near the Fermi energy, vicinity of $\textsf{a} =1$ is important, this point should be within the range of integration over $\textsf{a}$ in Eq.\ \eqref{eq:MF-UE}, i.e. $\textsf{a}_- < 1 < \textsf{a}_+$. The latter condition is fulfilled provided $4 t \sqrt{2/t_0} \gg \ln (1/\alpha)$.

\bibliography{biblio}

\begin{thebibliography}{58}
\bibitem [{els()}]{elsewhere}%
  \BibitemOpen
  \href@noop {} {}\bibinfo {note} {To be published elsewhere}\BibitemShut
  {NoStop}%
\end{thebibliography}

\end{document}